\documentclass[12pt]{article}
\usepackage{graphicx}
\begin{document}
%\baselineskip=1.5\baselineskip
\begin{center}
{\Large Quantum Cosmological Models }\\
\bigskip
{\bf D.H. Coule}\\
\bigskip
Institute of Cosmology and Gravitation,\\ University of
Portsmouth, Mercantile House, Hampshire Terrace, Portsmouth PO1
2EG.\\
\bigskip

\begin{abstract}

We contrast the initial condition requirements of various
contemporary  cosmological models including inflationary and
bouncing cosmologies.

Canonical quantization of general relativity is used, as a first
  approximation to full quantum gravity, to determine
  whether suitable initial conditions are present.

Various proposals such as Hartle-Hawking's ``no boundary '', or
Tunnelling boundary conditions are assessed on grounds of
naturalness and fine tuning. Alternatively a quiescent initial
state or an initial closed time-like curve  ``time machine''  is
considered. Possible extensions to brane models are also
addressed. Further ideas about universe creation from a
meta-universe are outlined.

Semiclassical and time asymmetry requirements of cosmology are
briefly discussed  and contrasted with the black hole final state
proposal.

 We compare the recent loop quantum cosmology
 of Bojowald and co-workers with these earlier schemes. A number of
possible difficulties and limitations are outlined.


\end{abstract}
\end{center}
\newpage
{\bf Contents}\\
1.0 Introduction\\
2.0 Classical Initial Conditions\\
\hspace*{5mm} 2.1 Initial conditions for Inflation\\
\hspace*{5mm} 2.2 Finite domain size for Inflation\\
\hspace*{5mm} 2.3 Inflation from previously non-inflationary
conditions\\
\hspace*{5mm} 2.4 Variable constant models\\
3.0 Quantum Cosmology\\
\hspace*{5mm} 3.1 Cosmological constant $\Lambda$ case\\
\hspace*{5mm} 3.2 Massless scalar field case\\
\hspace*{5mm} 3.3 Scalar potential $V(\phi)$ case\\
\hspace*{5mm} 3.4 Further Topological and Geometric aspects\\
\hspace*{5mm} 3.5 Arrow of time and classical description\\
\hspace*{5mm} 3.6 Brane quantum cosmology\\
\hspace*{5mm} 3.7 Universe from a quiescent or static state\\
\hspace*{5mm} 3.8 Black hole final state\\
4.0 Loop Quantum Cosmology\\
\hspace*{5mm} 4.1 Bouncing and Inflationary model? \\
\hspace*{5mm} 4.2 Modified Wheeler-De Witt equation\\
\hspace*{5mm} 4.3 Loop cosmology and super-inflation\\
\hspace*{5mm} 4.4 Is loop quantum cosmology unstable?\\
\hspace*{5mm} 4.5 Summary: loop and quantum cosmology\\
 5.0 Conclusions \vspace{2cm}

 {\bf 1.0 Introduction }

 In this review we intend to outline and compare various ideas
 about the origin of the universe as suggested by quantum
 cosmological reasoning. A number of earlier reviews are already
 available especially concerning technical issues [1-12]. We also
 consider the
 implications of possible discreteness of space, as suggested
 by loop quantum gravity - for extensive reviews see [13,14]. For other
 approaches and more general
 issues of quantum gravity see e.g. [15-25].

The recent history of cosmology is one of gradually  more extreme
matter sources being considered. Originally in standard big bang
(SBB) cosmology only ordinary matter like radiation or dust, that
satisfies the strong energy condition, was believed suitable. This
resulted in singularities being present and other so-called {\em
puzzles} of SBB cosmology.

A significant step was made when false vacuum states were
considered that violate the strong energy condition: so driving an
antigravity phase, now generally known as inflationary expansion -
see e.g. [26,27], while for definition of various energy
conditions see [28-30]. Further cosmological issues and
alternative models are given in [31-35].

 Recent
evidence suggests the universe is again undergoing inflationary
expansion although with perhaps a very small effective
cosmological constant, this and other possibilities are  reviewed
in [36]. However, in the early universe many of the original
puzzles for inflation have since been shown to be either
avoidable, solvable by other means or  anyway unsatisfactory
treated by inflation. For example, the {\em horizon problem}
[26,27] is something that is actually generated during the Planck
time [37], so a quantum gravity solution is really required:
perhaps by the presence of wormholes in the quantum foam at that
time e.g.[38]. Since thermodynamical quantities become ill-defined
at the Planck epoch cf.[39] it might be expected anyway that
initial homogeneity or smoothness could occur [40]. Another
drawback is that a  secondary {\em horizon problem} caused by
phase transitions proceeding after inflation could occur and so
spoil the earlier smoothness [24].

 The {\em flatness problem} [26,27] might also be solved
by quantum consideration during the Planck time [41], ( see p.501
in ref.[33] for a related solution )  or by choosing natural
values for arbitrary constants [42].\footnote {There is a Bayesian
argument [43] that also apparently solves the flatness problem,
but it assumes an invalid scale invariance property of General
Relativity [44].} This choice either given by quantum gravity
arguments or from purely classical reasoning.

Neither is the  presence of singularities in SBB cosmology
alleviated by inflation. Recent proofs suggest inflation also
cannot be extended indefinitely into the past [45]: essentially
because the flat de Sitter metric is then geodesically incomplete
[28] (to null rays) unless one makes topological identifications
that remove time orientability [46]. Earlier incompleteness
theorems required the weak energy condition be satisfied [45]. One
unorientable example of patching together two de Sitter universes
with time running oppositely is given in ref.[47], although this
idea has long been known [48]. The result can also be questioned
on the grounds that actual finite particle lifetimes never allows
timelike paths to actually become null cf.[49]. Recycling
behaviour might also help circumscribe   the geodesic
incompleteness results [50].

One major advantage of inflation is that it can give a large
universe even starting with natural Planck values for various
quantities: so an initial   ``small bang'' can be a sufficient
starting point. Without inflation  the energy density when
extrapolated back vastly exceed  the Planck energy density when
the universe was the Planck time old: the {\em Planck problem} of
SBB cosmology [51,26,27]. Another use of inflation is that Hawking
radiation can provide a mechanism for providing scale-invariant
perturbations for galaxy formation e.g. [52].

For a simple FRW universe the Friedmann equation is given by
[26-29]\footnote{ We use notations and units of the quoted
references, generally  Planck units.}
\begin{equation}
H^2+\frac{k}{a^2}=\frac{8\pi G}{3}\rho
\end{equation}

 For a perfect fluid with equation of state $
p=(\gamma-1)\rho$ the energy density $\rho$ scales as
$a^{-3\gamma}$. The strong energy condition is violated for $0\leq
\gamma <2/3$. Such an example is that of a scalar field with
potential $V(\phi)$ dominating over the kinetic energy
$\dot{\phi}^2$ . In general, while keeping $\rho$ positive, and
since
\begin{equation}
\gamma=\frac{2\dot{\phi}^2}{\dot{\phi}^2+V(\phi)}
\end{equation}
one can obtain more extreme violation of the weak e.g. ($ \gamma
<0$) or dominant energy conditions ($\gamma>2$)  by choosing, in
turn, either a negative kinetic term or negative potential. The
negative $\gamma$ example, from a ``wrong sign''  kinetic term,
causes super or pole-law inflation: the scale factor $a$ goes as
$a\sim t^n$ with $n<0$. If the present expansion is of this form
it has been dubbed a phantom matter source proceeding towards a
``big-rip'' singularity [53]. Although such models were earlier
simply described as a ``whimper expanding to a big bang''
e.g.[31]. Numerous examples of super-inflation are known such as
in Brans-Dicke gravity e.g.[32], with parameter $\omega<-3/2$ or
with coupled dilaton-axion scalar fields [54]. Super-inflation
suffers from a number of problems stemming from a growing Hubble
parameter giving a blue spectrum of perturbations and the presence
of a future singularity - so was generally discounted as a
suitable form of inflation  for the early universe [55].

If the present universe is undergoing such a rapid expansion it is
difficult to see how a phantom scalar field is compatible with an
earlier inflationary phase in the universe. The required ``wrong
sign''  kinetic term should have been diluted or else it would
have dominated over the earlier inflationary source. Having
``wrong sign'' spatial gradient terms is also likely to cause
instability [56]. Another possibility  is to add various new terms
to the gravitational action
 and so  allow super-inflation today e.g.[57], although extra non-analytic
 Ricci scalar terms generally only give power-law inflation  i.e. $a\sim t^n$
 with now $\infty>n>1$ [58].

 The ekpyrotic model [59] is  a
recent attempt at producing a cyclic universe inspired by the
supergravity limit of string theory. During the contraction phase
a negative potential can dominate to produce an effective
``super-stiff'' equation of state $\gamma>>2$ which is said to
produce an ultralocal phase.

Another  extreme case is to take $\rho$ negative, which is  a more
drastic  violation of the weak energy condition. One can now
produce a bounce for arbitrary curvature $k$ unlike for the closed
case which can bounce by violation of the strong energy condition
alone [60]. In general a FRW bouncing model is described by an
equation of the form
\begin{equation}
H^2=\frac{A}{a^n}-\frac{B}{a^m}
\end{equation}
$A,B$ arbitrary positive constants and provided $m>n$, a bounce
proceeds. Note that the stiffer (larger $\gamma$)  matter
component requires the minus sign. For the closed universe the
curvature plays this role i.e. $m=2$, so only the strong energy
condition has to be violated i.e. $0\leq n<2$ for a bounce to
proceed.

 Such bouncing models tend to suffer from
rapidly growing classical perturbations since the usual Jeans
instability is no longer damped by rapid expansion [61,62].
Further examples of bounces  are reviewed in ref.[63], while the
eventual dominance  of any (small) cosmological constant was shown
in ref.[64]. The related quasi-steady-state model [65] has an
underlying de Sitter expansion with superimposed oscillations: the
model now has a ``no hair'' property that inhomogeneity is
suppressed [66]. Although the bouncing models can avoid a possible
singularity they still require suitable initial conditions, albeit
at an arbitrary distant past. If such models collapse as much as
they expand it will tend to negate any advantage of having an
inflationary phase. We later will use this to restrict the
usefulness of combining bouncing behaviour with subsequent
inflationary expansion.

We expect a quantum description to more realistically describe the
 early universe with probably high
curvature regions where quantum gravity would  actually dominate.
As a first approximation one can attempt to quantize Einstein's
field equations and obtain quantum cosmology using predominantly
the Wheeler-De Witt (WDW) equation [67,68]. Ultimately, the true
theory for the forces of nature will automatically include quantum
aspects and such an approximation would be superseded. It is
sometimes said that quantum mechanics will have to forgo some of
its principles for this unification, see e.g. [22] and its
validity might cease at suitable large ``size'' [24,25] or
complexity [69]. However, in any case quantum mechanics alone does
not necessarily regularize singularities. Recall the
Schr$\ddot{o}$dinger equation for the hydrogen atom in $N$ spatial
dimensions. The total energy is given by
\begin{equation}
E\sim \frac{h^2}{2mr^2}-\frac{e^2}{r^{N-2}}
\end{equation}
where the Heisenberg uncertainty principle is used to estimate the
kinetic energy resisting localization. Following the presentation
in ref.[33], when  $N>5$ the potential is too divergent and there
is no minimum energy level [70]. Using more rigorous arguments
that include special relativistic effects this result can be
strengthened such that no stable bound exists for $N>3$ [71]. So
the quantization alone does not prevent the $(N>3)$ hydrogen atom
being singular. Similar problems are expected to occur in quantum
gravity if the potential term is initially too divergent.

Even if the singularity becomes regularized it is still perhaps a
mixed blessing. If there is no longer any breakdown in the
evolution one then requires an explanation as to what instead
happens. If there is some automatic limiting curvature then
perhaps a de Sitter phase occurs near the singularity [72] or
simply some entire new universe results [73]. If one simply
reenters the same universe, or passes into a parallel one, then we
no longer have any natural starting point to the universe. Or
perhaps the singularity then matches to some ubiquitous  natural
state.

One widely used notion in  quantum cosmology is to assume such a
state of ``nothing'' or zero size from which universes can
initially tunnel [1,2]. More recently a possible meta-universe has
instead been suggested  as a potential source of universes
[74-76]. The idea is that a, usually disfavoured, low entropy
initial state is then possible   by tunnelling like behaviour from
the meta-universe. This unfortunately then pushes problems of
naturalness into this pre-existing state, which then itself
requires further explanation. If this meta-universe is already
perhaps  at a maximum entropy state then fluctuations could just
as well surpass the holography bound [77] causing singularities.
If the meta-universe is continually expanding then again it will
likely suffer from related geodesic incompleteness theorems
cf.[75,45].

Alternatively a continual cyclic evolution of a single universe,
through possible regulated singularities, as in the ekpyrotic
example, might be a less drastic solution if one wishes to
entertain  the notion of an eternal state. But there are still
issues as to whether a ``heat death'' or maximum entropy state
should rather be expected [78].

 If the present universe is
undergoing super-inflation with a growing Hubble parameter it
might eventually reach a quantum gravity singular state: perhaps
eventually matching to a more  moderate high energy de Sitter
expansion. This state might then spawn new universes in a cyclic
like manner. Whether such a scheme is possible is presently a
topic of some interest but again would appear profligate in terms
of growing space and overall entropy production.  The geodesic
incompleteness results might again be alleviated by giving all
matter only a finite lifetime.

We can consider different examples  of how quantum effects might
play a role in the  quantum gravity, or early, phase of the
universe.\\
$\bullet$ {\em Different matter sources.}

Quantization  will only have significant differences when the
classical evolution reaches regions of small action. This is the
most studied case where different matter sources are investigated
and quantum tunnelling of spacetime itself considered when the
total action is small. One problem is that action is rather
  more ambiguous when gravity is included. \\
$\bullet$ {\em Alter time.}

 Time might have more unusual
properties during an quantum gravitational epoch. One possible
example is that closed timelike curves [28,30] might occur during
a quantum gravitational epoch. Adapting quantum mechanics to this
example of a ``time machine'' is rather involved before one even
considers
including high curvatures. \\
$\bullet$ {\em Alter space.}

 This approach stems from the idea
that space itself should be quantized so that it occurs in
discrete packets or ``cells'' [14]. Eventually at larger distances
the cells merge together to approximate a continuum.  Recently
loop quantum cosmology [13] has been developed and is said to
allow matter to take different properties when this granularity of
space is apparent. One difficulty we will address is that absolute
distance can be a rather nebulous quantity especially during rapid
cosmological expansion.

Just as the hydrogen atom is dependent upon the form of the
potential we first need to consider a suitable cosmological model
before any quantum aspects might further be introduced.

 {\bf 2.0 Classical Initial conditions}

 There are essentially two approaches one can take for the initial
 conditions: chaotic or orderly. In the earlier chaotic cosmology
 program, dissipation or quantum particle effects were supposed to
 smooth the universe - see e.g.[33]. In practice, such processes can only work if
 the initial anisotropy is constrained.  One could choose
 unwanted conditions in the universe today and evolve backwards
with the deterministic Einstein's equations to obtain initial
conditions that are not-consistent with our universe [79]. The
celebrated ``no-go'' theorem of Collins and Hawking [80] also
shows that only a measure zero of universes become isotropic at
late time.

Inflationary cosmology tried essentially the same program but with
a cosmological constant $\Lambda$ or scalar potential $V(\phi)$
playing the role of the smoothing process. The hope that ``no
hair'' of the chaotic initial conditions would remain can be
contradicted by choosing large initial inhomogeneities - see [81]
for a review. Likewise one could evolve backwards unwanted
conditions through a finite amount of inflationary expansion
[79,82,83]. The initial conditions therefore require some
restrictions in order for a smooth FRW universe to be eventually
produced. One can obtain a similar Collins-Hawking result for an
inflationary convex potential $V(\phi)$ with zero minimum [84] or
(flat) exponential potential provided some inhomogeneity is
initially present [85]. In some general sense the scalar field is
less robust at removing hair than the pure cosmological constant,
e.g. [86].

It would therefore seem that inflation alone does not entirely
explain the obeyance of the {\em cosmological principle} e.g.[34].
Although the Collins-Hawking requirements are probably
unnecessarily  strict, since smoothing for finite future time and
shear remaining bounded $\sigma/H\rightarrow $ constant [33,79],
would suffice in practice to create a temporary observance of the
{\em principle}. However counter examples are known so some
restrictions on initial conditions are required if inflation is to
actually produce a FRW universe from arbitrary initial conditions.
One question is whether a (quantum) principle alone can provide
this restriction so allowing inflation some chance of success?

The ekpyrotic universe also attempts to smooth space  by using
both a low-energy driven inflationary phase and a ``super-stiff''
$\gamma>2$ phase during a subsequent contraction [87]. This
combination might then  have a more robust  ``no hair'' property
than inflation alone.  But again  this could also be  overwhelmed
by choosing irregular staring conditions: massive black holes for
example that are not ripped apart by inflation or have time to
evaporate [78]. The argument now for reducing anisotropy is that
as the bounce proceeds, the super-stiff matter scales as $\sim a
^{-3\gamma}$ while the shear $\sigma^2\sim a^{-6}$. So for
$\gamma>2$ the shear remains subdominant and the universe is
forced isotropic [87,88]. However it is also important that the
individual components should remain below Planck values if the
model is to be described by the low energy supergravity limit of
string theory. Without inflation around the conventional big bang
point one requires energy densities vastly bigger than Planck
values; the {\em Planck problem} of conventional cosmology [51]. A
similar problem occurs in another recent attempt at avoiding
inflation during the conventional big bang type evolution [89].

Inflation must also restrict the nature of space at the Planck
scale: length scales are being blown up from below the Planck
length to macroscopic size. If space had a fractal like structure
this complexity would simply be extended to large size by
inflationary expansion [90,79]. This problem is particularly
severe for perpetually expanding models such as, from the higher
dimensional perspective, the ekpyrotic scenario. Since it uses low
energy string theory as a starting point it is eventually
inconsistent that all length scales within the universe have come
from below the Planck or string length of previous cycles [78].

The present string paradigm is that branes are present in higher
dimensional bulk spaces - see e.g. [91] for a review. Our usual
matter is confined to 4 dimensional spacetime while gravity, and
some types of exotic  matter, extends freely over the extra
dimensions [92]. The branes are believed to undergo cosmological
evolution presumably from some finite time in the past. But  so
far these models have been done with highly symmetric bulk spaces
particularly Anti-de Sitter (AdS) [28] space with maximal
symmetry. This is a strict departure from the idea of random
initial conditions to spaces that obey the unchanging {\em perfect
cosmological principle} e.g.[34]. Some attempts have been made to
first have inflation within the bulk space but this is rather
contrived if a negative cosmological constant is the true ground
state [93]. Also inflation, driven by matter that introduces a
preferred time cf.[94], does not provide such idealized symmetry,
as previously discussed, at best it can only achieve the {\em
cosmological principle} for finite times. Apart from simply
passing to a higher dimensional cosmology, that requires
explanation,  one requires the initial conditions or processes
that can produced the original smooth bulk space.

To more closely match the bulk space one can attempt to  make the
brane eternal by achieving a bounce. One example was by taking a
Reissner-Nordstr$\ddot{o}$m AdS bulk which can modify the
Friedmann equation to [95,96]
\begin{equation}
H^2+\frac{1}{a^2}=\frac{8\pi G}{3}\left(
\rho+\frac{\rho^2}{2\lambda}\right )
+\frac{M}{a^4}-\frac{Q^2}{a^6}
\end{equation}
where $\lambda$ is the brane tension and M and Q represent the
mass and charge of the bulk space. However this charged black hole
appears unstable to perturbations since the bounce occurs within
the outer event horizon [97,98]. Related work has suggested only
weak singularities occur because of related ``blue sheet'' effects
[99], but this probably required restricted initial conditions.
The difficulty is that one is effectively  trying to violate the
weak energy condition and this often leads to instabilities. Note
also that by choosing the wrong sign of the brane tension in
eq.(5) you can also get a bounce without the bulk being charged
[100] cf. eq.(3). Being on a negative tension brane is probably
not consistent with other solar system constraints, and would also
require more than a single brane to stabilize [101]. One can also
obtain a similar bounce by making the extra dimension  time-like
[102], although this scenario seems likely to be restricted by
observations limits cf.[103].

We can contrast this brane approach with the  earlier pre-big bang
model [104] that used the low energy limit of string theory with a
dilaton term - reviewed in [105]. The extra dimensions are now
assumed tightly compactified. This model is taken to start at time
$t=-\infty$ before approaching a singularity at, say, time $t=0$.
However this model is not inflationary in the Einstein frame, so
still having a {\em Planck problem} [106]. Also having no ``no
hair'' like suppression requires fine tuning of initial
conditions. A suggested initial state of plane gravitational and
dilatonic waves is justified as having zero gravitational entropy
``asymptotic past triviality'' [107]. These waves after
gravitationally interacting are expected to form a space-like
singularity into the future which plays the role of the usual big
bang singularity

 During any collapsing phase quantum fluctuations can be produced
by Hawking radiation although one has to have the correct
contraction rate to get a scale invariant spectrum: dust
$\gamma=1$ [108,109] or super-stiff equation of state $\gamma>>2$
as in the ekpyrotic case [59,87]. Firstly it is not clear that
these quantum fluctuations will adequately decohere into classical
perturbations cf.[110].  Also quantum fluctuations require the
presence of an event horizon, this coming from the time reversal
of a particle horizon [34] which means the strong energy condition
now has to be satisfied for quantum fluctuations to be present: so
now being the opposite to the usual inflationary case. However,
because if the impending singularity is to be avoided, the energy
conditions have eventually to be violated. This, strictly
speaking, contradicts the requirements for the production of
Hawking radiation during the collapsing phase. There are similar
uncertainties when considering the Unruh radiation within a
Rindler wedge (see e.g.[111])  when acceleration proceeds for only
a finite time cf.[112].

Obtaining the weak energy violating source that turns around the
pre-big bang has proved elusive, although certain higher order
gravitational action corrections terms from string theory are
possible. But requiring, for example, the wrong
 sign for the  Gauss-Bonnet term  can cause
 other instabilities e.g.[113].

 In summary most emphasis in cosmology has been on assuming random initial
 conditions and then hoping some process will  smooth things gradually
 towards a FRW like universe. In brane cosmology highly symmetric
 models have been assumed {\em ab initio}. Can quantum
 considerations suggest which of these complementary cases  is most reasonable?

{\bf 2.1 Initial conditions for Inflation}

We can consider a simple example of a scalar field with energy
density $\rho=\dot{\phi}^2+V(\phi)$  within a FRW universe  to
demonstrate some of the issues involved.  Such a model can display
chaotic inflation [114,115].  Unlike a simple cosmological
constant which only has a de Sitter solution the scalar field has
both inflationary and non inflationary solutions. In order to
classify which is most probable Belinsky and coworkers [1196-119]
considered a simple equipartition at the Planck boundary $\rho\sim
M^4_{pl}$. This was motivated by the, admittedly  rough, argument
that the uncertainty principle $\Delta E \Delta t\ \geq 1$ applied
during the Planck time would give an uncertainty in the energy
$E\sim$ Planck value [116,27]. This argument also assumes that the
universe be closed so that the total energy is zero: positive
matter energy density being cancelled by negative gravitational
energy. Otherwise the uncertainty in energy would have to be paid
back before the scalar field could take large classical values: or
the non-zero total energy would have to be justified further.

 This sort of analysis for a massive scalar
field $V(\phi)=1/2m^2\phi^2$ gives the probability of sufficient
inflation $\sim70$ e-foldings around $\sim(1-m/M_{pl})$ [116] so
almost certain for mass $m<<1$. For the correct size of
fluctuations to be produced requires $m\sim 10^{-5}$ [26,27]. The
slight chance of non-inflation is when the initial kinetic energy
$\dot{\phi}^2$ is large  $\sim 1$ in Planck units [116].

A more rigorous approach is to use a canonical measure [120-123]
\begin{equation}
 \omega =-dp_a \wedge da +dp_{\phi} \wedge d\phi
 \end{equation}

 where the canonical momenta are
 \begin{equation}
 p_{\phi}=a^3\dot{\phi}\;\;\;\; p_a=-3a\dot{a}
 \end{equation}

at fixed scale factor $a=a_0$ this measure simplifies to [42,122]
\begin{equation}
\omega=a_0^3d\dot{\phi}\wedge d\phi
\end{equation}

Because this measure is peaked at large energy densities it can
solve the flatness problem i.e. that $\Omega=1$ regardless of
inflation [42]. Although if the potential is bounded above,
$\Omega$ can again become arbitrary [121,122]. Such a bounded
potential might be expected from a conformal anomaly or higher
derivative $R^2$ like terms in the action [122].

The canonical measure can be shown to give an ambiguous
probability of whether inflation occurs: there are infinite
numbers of both inflationary and non-inflationary solutions [42].
Although this sort of ambiguity might be rectified if some input
from quantum cosmology could, for example, determine if the
universe started small or with a certain energy density [122]. One
can also investigate the probability of bounces occurring. In
simple models most collapsing universes have only finite measure
of bouncing compared to infinite measure of singular solutions
[121,125]. We are therefore not sure that one can say that
deflation is as likely as inflation with this measure cf.[123].
There is also an extremely small measure for perpetually
oscillating solutions: both periodic and non-periodic [126,127].
If bounces are imposed, perhaps by a limiting curvature cf.[72],
then the model can eventually become inflationary by means of a
growing value of $\phi$, provided the scalar field can oscillate
around a minimum [128].

Once the scale factor is fixed one can obtain a finite total
measure and obtain the probability of inflation [42,120,121]. It
depends as expected with the chosen energy density. For the
massive scalar field case the fraction of inflationary solutions
$f_{I}$ , for at least $z$ e-foldings of inflation is [42]
\begin{equation}
f_{I} \sim 1-\frac{m\ln (1+z)}{\rho^{1/2}}
\end{equation}

Taking $m\sim 10^{-5}$ and requiring 100 e-foldings of inflation
gives $f_{I}\sim 1-10^{-5}$ for an assumed energy density $\rho
\sim $ Planck value. The above expression gives that as the
initial value of $\rho$ is reduced $f_{I}\rightarrow 0$. Although
a more detailed calculation shows $f_{I}$ actually tends to a very
small finite value as $\rho$ is taken to zero [42]. However, if
anisotropy or other inhomogeneous degrees of freedom are included
then $f_{I}$ would also approach zero as $\rho\rightarrow 0$
[129]. In order to give a high probability of inflation therefore
requires that the energy density should be large with this measure
once the initial scale factor is fixed. We later will consider how
 loop quantum effects attempts to remedy
 this prediction and obtain inflation even for small
 initial values of the energy density.

 {\bf 2.2 Finite domain size for Inflation}

So far this analysis assumes the scalar fields are homogeneous
throughout the universe. But if only some finite  domain is to
inflate  then this
 homogeneous region must be of
 sufficient size. Recall that the scalar field behaves as a
 negative pressure and any outside positive or zero pressure will
 wish to equalize the disparity   by ``rushing in'' to cancel
 any pressure differences. Assuming this
 equalization can proceed at the speed of light one finds that the
 homogeneous domain must be of a size greater than $\sim H^{-1}$
 [81]. This can be alleviated in some topologically  non-trivial universes cf. [130]. Also
 topological driven inflation [131] using defects, such as monopoles,  can possibly  weaken
 this requirement due to topological charge conservation [132];
 but we
 leave aside these possible complications.
 Once such a suitable domains occur then inflation will proceed.

 Because of a combination of  the fluctuations
 that are generated during inflation and
 the finite horizon size $\sim H^{-1}$ in de Sitter
 space inflation never stops entirely once started, provided a
 requirement on the size of the scalar field $\phi>\phi_*$ is
 satisfied
 [115]. In general the potential must satisfy
 $V(\phi)^{3/2}>dV(\phi)/d\phi$ for eternal inflation [27].

  If the initial domain
 does not have such  a sufficiently large scalar field value one can hope that
 a quantum ``instanton'' effect can produce a domain with sufficiently large
 scalar field provided there is a sufficient  number $\sim \exp (1/V(\phi))$ of  Hubble volumes: so that an
especially large fluctuation will occur somewhere [133,134]. Note
that this would require an astronomical $\sim \exp(10^{120})$
Hubble volumes for the apparently present inflationary state of
the universe to actually produce a suitable eternal domain.

  However,  there are a number of uncertain aspects [135] about this eternal inflation mechanism.
  For example whether the fluctuations of matter
   transfer to the geometric left hand side of Einstein's equation, or whether
   the black hole calculation entirely extends to the de Sitter case e.g.[136].
   Extra dimensions, perhaps the bulk space in brane models might
   regulate  such a mechanism cf. [137]. Other criticisms are given in
   ref.[138].

    But even if the eternal mechanism proceeds
   you can argue that the total produced number of new inflationary
   domains  is less than countable infinity $\aleph_0$
   for a finite time of
   eternal universe production [135]. It does not turn an
    initially finite universe into an infinite one. It therefore
   is not clear that this mechanism alone can overcome, for example,  the previous ambiguity
   in the  classical measure that an infinity of both inflationary and
   non-inflationary solutions are possible.
   You might as well  start
   with an initially infinite universe, like the flat FRW one, and put
   down random conditions - but this
   seems a extravagant way of trying to counteract an infinity of
   unwanted solutions.

   What is required is a definite prediction that inflation will
   occur - the idea that once inflation starts it will never finish does
   not supersede this requirement.

 {\bf 2.3 Inflation from
previously non-inflationary conditions}

 What about the naturalness of inflation in models that do not
inflate initially from their inception.
  Singularities seem a general consequence of producing conditions
  that give inflationary behaviour. The method of producing a
  ``universe in the laboratory'' was required to expand so rapidly to
  avoid re-collapse that a singularity would be present [139-141]. This can
  be seen clearer by noting that in a FRW universe, regions of size
  bigger than the so-called apparent horizon $\sim \rho^{-1/2}$,
   have a necessary singularity -see page 353 in ref.[28]
   where such a quantity
   is called the {\em Schwarzschild length} of matter density $\rho$.
  But this size, $\sim H^{-1}$ for
   the flat $k=0$ case, is the minimum required to isolate an inflationary
  patch from its surroundings for sufficient time to start
  inflating,  ignoring possible topological counterexamples.
   Now requiring this initial patch  size to be
  larger than the apparent horizon size has been emphasized [142]
  as a  problem in setting up inflationary conditions.
   However, depending on the matter source it need not strictly
  violate causality which is rather determined by the particle horizon: the
  distance light travels from the beginning of the universe. The
  large initial patch size does though imply that a singularity is present [7].

 If we consider further  the nature of the {\em horizon problem}. It
 occurs because the the particle horizon size, defined as
\begin{equation}
r=c\int_{0}^{t} \frac {dt}{a(t)}
\end{equation}
is finite, see eg.[26,27]. The horizon proper distance $R$ is
 this quantity $r$
multiplied by the scale factor i.e. $R=a*r$. For any strong energy
satisfying matter source this quantity $R$ grows linearly with
time. But in SSB cosmology the rate of change of the scale factor,
given by $a\sim t^p$ and $0< p<1$, grows increasingly rapidly as
$t\rightarrow 0$. The horizon cannot keep pace with the scale
factor ``velocity'' $\dot{a}\sim 1/t^{1-p}$. But  this only occurs
for times below unity $0<t<1$. If the horizon problem was already
solved at the Planck time $t_{pl}$
 it would remain  permanently solved during the ensuing evolution [37].
  Note
 also that in models that inflate from their inception the usual
space-like singularity of the FRW universe becomes  null like when
$p>1$ - so any horizon problem is absent, see eg.[28,143].

The main idea of inflation is to take an initial domain of size
less than the corresponding particle horizon size and allow it to
expand greatly to encompass our universe. Let us see how this
requirement can be satisfied. Most of the relevant quantities have
already been obtained in the context of the holography conjecture
[144,145].

 For this purpose a  useful form
of the FRW  metric is
 \begin{equation}
 ds^2=a^2(\eta)\left ( -d\eta^2 +d\chi^2 + f^2(\chi) d \Omega^2 \right )
 \end{equation}
 where $f(\chi)=\sinh \chi\; , \chi\; , \sin\chi $ , corresponding
 to open, flat and closed universes respectively. We can define a number of important quantities.
 The {\em Hubble horizon} is defined by
 \begin{equation}
 r_H=H^{-1}
 \end{equation}
  The {\em particle horizon}, or the distance travelled by
 light from the initial moment of the universe, is simply,
 \begin{equation}
 \chi_{PH}=\eta
 \end{equation}
 for this metric. The {\em apparent horizon} is given by [144]
 \begin{equation}
 \chi_{AH}=\frac{1}{\sqrt{H^2+k/a^2}}\Rightarrow
 \frac{1}{\sqrt{\rho}}
 \end{equation}
Roughly speaking light rays beyond the apparent horizon are seen
to move away from the origin, a so-called anti-trapped behaviour
e.g. [28,29]. Note that in the flat case $k=0$ the apparent
horizon and Hubble horizon coincide.

 In ref.[142] it is argued that the initial inflationary
patch must have sufficient size $x$ that it reaches the
anti-trapped surface i.e. $x>\chi_{AH}$. Otherwise the weak energy
condition is violated for light rays that could otherwise enter
the inflating region from normal or trapped regions. For a perfect
fluid  $p=(\gamma-1)\rho$, the apparent horizon has the following
time dependence [144,145]
\begin{equation}
\chi_{AH}= \frac{d\gamma-2}{2} \eta
\end{equation}
with $d$ the number of space dimension. However the causal
particle horizon has a different time dependence simply
$\chi_{PH}=\eta$ so the condition
\begin{equation}
\chi_{AH}<x<\chi_{PH}
\end{equation}
can be satisfied for
\begin{equation}
\frac{d\gamma-2}{2}<1 \;\; \stackrel{d=3}{\rightarrow} \;\;
\gamma< 4/3
\end{equation}
This does exclude the case of radiation ($\gamma=4/3$) or stiffer
equations of state. But if $\gamma$ was gradually reducing before
inflation occurred this causal constraint can be satisfied. The
condition can be thought of as saying the effective value of
$\gamma$ cannot switch suddenly but rather must fall below $4/3$
for sufficient time to allow the causal or particle horizon to be
larger than the apparent horizon. This result is independent of
whether curvature is present.

 In the closed case
only during the expansion phase is an anti-trapped surface present
- cf. Fig.(4) in ref.[145]. This means that producing inflation,
within a larger domain, to avoid an impending big crunch
singularity during a collapsing phase will violate also the weak
energy condition.
 Now it is the case that needing $x> \chi_{AH}$ is difficult to
justify in terms of particle physics processes, but if this patch
could be smaller than $\chi_{AH}$ one could avoid the singularity
in a FRW universe since the matter would be insufficient to
converge the light rays into the past. See chapter 10 in ref.[28]
for a proof of this argument. So allowing an initial domain of
size $x<\chi_{AH}$ to inflate, would have allowed singularities to
be expunged from this cosmology: the result that this cannot be
done without violating the weak energy condition is therefore
consistent with  studies of eternal inflation that singularities
have to be present when the model is continued into the past [45].
Note that the more general geodesic incompleteness results show
that even violation of the weak energy condition does not prevent
past geodesics leaving the metric, although not necessarily
possessing curvature singularities.

There are some alternative metrics with non-singular solutions,
but like Minkowski space they do not have anti-trapped regions
[143,146]. Achieving inflation in such spaces would likewise
require the violation of the weak energy condition.

However such violations of the energy conditions are possible with
quantum fluctuations of the vacuum. Although  the magnitude and
duration of such violations are constrained by so called quantum
inequalities cf.[147]. Universe creation in the laboratory by
quantum tunnelling [148,149], or starting from Minkowski space or
at high temperatures has been considered [150]. A black hole is
produced in the laboratory that eventually evaporates allowing the
new universe to disconnect from the original one: it does not
simply supplant it [148-150,50]. The process generally is enhanced
for larger coupling constants e.g. $m$ or $\lambda$ , and with
temperatures approaching Planck values [150]. Topological
inflation also produces charged black holes to observers outside
the defect [131]. This black hole does not evaporate though,
because of charge conservation, and any new inflationary universe
remains connected by a wormhole to the original space.

A possible meta-universe could  then take the place of the
laboratory and cause, by Poincar$\acute{e}$ recurrence, universes
to automatically branch off by quantum effects [74-76] , see also
Albrecht in ref.[23]. Although the probability is remote, if there
is infinite space or time, then such events will eventually occur.
This is not entirely  a satisfactory explanation since the first
cause is then simply switched into understanding the origin of the
meta-universe.

 {\bf 2.4  Variable
constant models}

 Another possible solution to the horizon problem is to
 postulate that the various constants, particularly the speed of light $c$, could take
 different values during the early universe [151], for a review see [152]. This alone is
 not too helpful since a space-like singularity cannot be crossed
 by any finite value of $c$ and a higher $c$ just means one has to
 go further back in time to see an equivalent horizon problem [153].
 There is also a causality problem, of sorts, as to why $c$ can
 change simultaneously over the whole universe and constantly stay equal
 throughout, once the value of $c$ has started to reduce and causal contact
 lost. This
 behaviour for $c$ really has to be pre-programmed in the universe from
 its conception: so the horizon problem appears
  simply switched into how constants ``know how to vary'': a
  compression of information is absent.

  Changing such constants also tends to  suppress any quantum
gravitational epoch at the beginning of the universe [153]. The
gravitational action at fixed time, for say a radiation dominated
universe, scales as  $S\propto c^3/G$ so increasing for bigger
$c$. However, this quantum epoch can surface at a later time,
which must be pushed sufficiently far into the future. In this
regard these model have some similarity with the pre-big bang
phase which also starts in  a classical state and tend towards a
quantum gravitational singular region - that becomes the start of
the big bang phase. This makes such variable constant models more
difficult to conceive of by quantum creation schemes if initially
we expect a quantum gravity phase with some small action. However,
quantum cosmology at present does not explain why the various
constants take their actual values and so they must be imposed by
fiat.

{\bf 3.0 Quantum Cosmology}

We now wish to consider how quantum considerations can suggest
ways of providing initial conditions for classical evolution. We
are here interested in conceptual and cosmological applications so
we will avoid technical concerns as much as possible.

In the general case there is the Wheeler-DeWitt equation $H\Psi=0$
and momentum constraint $H_i\Psi=0$
 equations [67,68]. Such an approach corresponds with Einstein's field equations
 [154]. A $3+1$ decomposition of spacetime is further assumed to
 define a Hamiltonian
 [1-3] although eventually this might be relaxed cf.[155].
  In mini-superspace models the momentum constraint
 equations are trivially satisfied and the WDW
 can be written with  kinetic and potential $U$ pieces [1-3].
\begin{equation}
\left (-\frac{1}{2}\nabla ^2+U(a,\phi)  \right ) \Psi(a,\phi)=0
\end{equation}
where we consider a FRW model with scale factor $a$ and  scalar
field $\phi$ matter source. The WDW equation is formally
independent of time, since time is a property defined within the
universe [67]. However, in  simplified FRW like, or fixed
background models one can use, for example, the scale factor $a$
as a preferred time variable [1,12,19]. Although conventional
quantum mechanics requires a notion of time with path-integral
methods a generalized quantum mechanics might also be used that
does not require such a time parameter [156].

Some important early developments are given in refs.[157-165,41].
 A general idea is to start the
universe with a small, in units of $\hbar$, action so that quantum
effects can be expected to be relevant. As emphasized by Zeldovich
[51] such a low action state necessarily requires an inflationary
phase in order to drive a  ``small bang'' universe to have
correspondingly large amounts of energy and action: so eventually
simulating  the usual big bang quantities.

 {\bf 3.1 Cosmological constant $\Lambda$ case}

 For the archetypal
example consider a closed universe with a cosmological constant
$\Lambda$. The WDW equation simplifies to, [166-170]
\begin{equation}
\left ( \frac{d^2}{da^2}+\frac{p}{a} \frac{d}{da} -U \right )
\Psi(a) =0
\end{equation}
with $p$ a factor ordering ambiguity and  $U$ the WDW potential
is
\begin{equation}
U=a^2-\Lambda a^4
\end{equation}
see Fig.(1).
\begin{figure}
\begin{center}
\includegraphics[width=6cm]{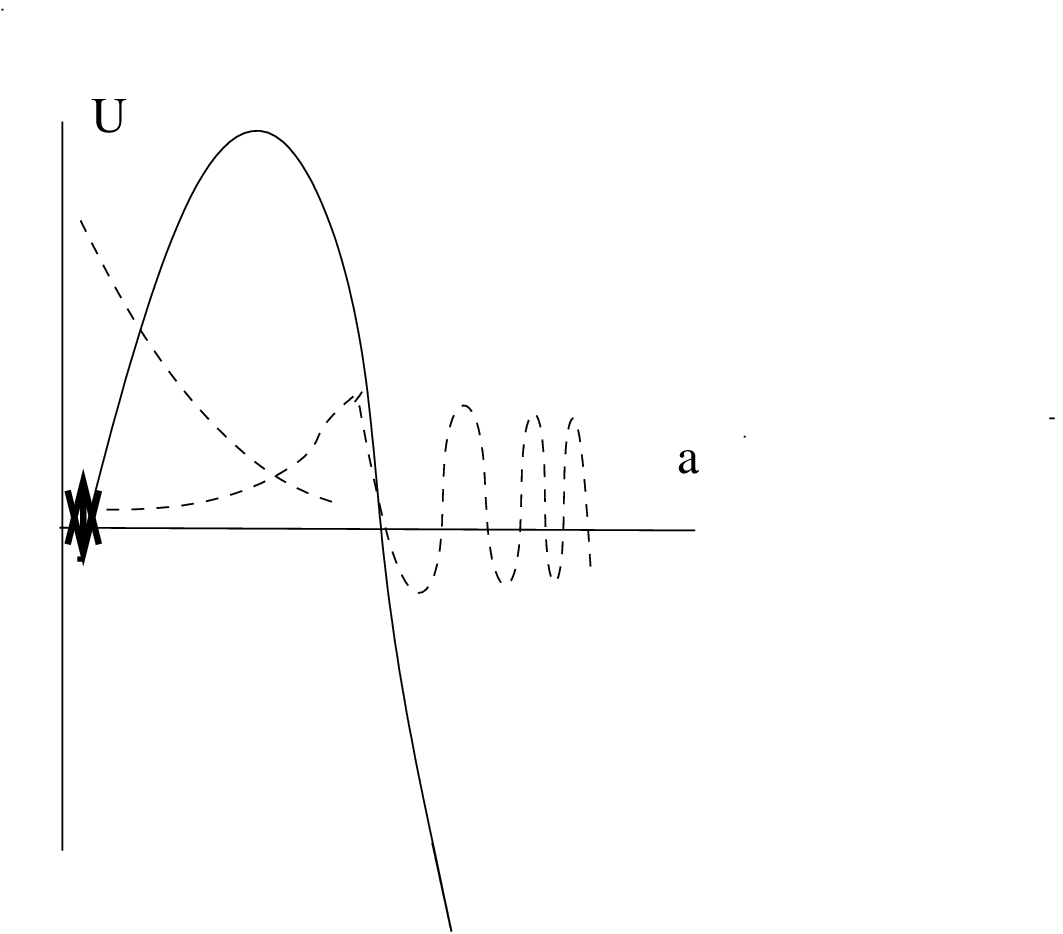}
\caption{The WDW potential (solid line) for the de Sitter example.
In the forbidden region $(U>0)$  exponentially growing and
decaying modes are present. Quantum uncertainty might  allow a
small classical universe to develop around $a=0$. }
\end{center}
\end{figure}
 For $U>0$ there is a
forbidden region: recall the classical de Sitter solution
$a=\Lambda^{-1/2}\cosh (\Lambda^{-1/2} t) $ that bounces before
reaching the origin. But instead of taking the classical
collapsing evolution from time $t=-\infty$ to $t=0$ it is
envisioned that the universe can tunnel  from $a=0$ or ``nothing''
to the smallest classically allowed value $a=\Lambda^{-1/2}$.

 For
the case $p=-1$ the WDW equation can be solved in terms of Airy
functions - see e.g.[167].
 The presence of a forbidden
region allows both exponentially growing and decaying solutions so
there can be large differences in behaviour e.g.[2,3]. The
solutions are sketched in Fig.(1).

  One can
also work with the gravitational action [29,171].

\begin{equation}
S=\int d^4x\sqrt{g} \left ( R - \cal{L_M} \right ) -2\int
d^3\sqrt{h} K
\end{equation}
where $\cal{L_M}$ is the matter component. One generally needs to
work with compact spaces to create a finite action or else add
further boundary counter-terms [172]. The wavefunction is given by
\begin{equation}
\Psi(h,\Lambda) =\int [g] \exp(i S)
\end{equation}
Formally the integration is over all geometries and topologies.
Hartle and Hawking  consider a rotation to Euclidean space
$t\rightarrow i\tau$ to improve the regularization properties
[165]. They further integrate over Euclidean manifolds with one
boundary, specified by the 3-metric h.\footnote{ It has been
suggested that one can obtain similar results  working with purely
Lorentzian metrics [173].} To semi-classical order the
Hartle-Hawking wavefunction is
\begin{equation}
\Psi_{HH}\sim \exp (-S)
\end{equation}
 We assume a number of technical difficulties can be resolved such
 as dealing with conformal modes that can set $S\rightarrow
 -\infty$ [174]. This wavefunction is considered to be a
 generalization of the notion of vacuum state to closed
 cosmological spacetimes [165,1-3].

The Euclidean action for a simple cosmological constant model is
given by [175-177]
\begin{equation}
S=-\Lambda V_4
\end{equation}
where $V_4$ is the 4-volume of the space. For positive $\Lambda$
the four-sphere $S^4$ has the largest volume with
$V_4=\Lambda^{-2}$ so the action is negative $S=-1/\Lambda$.
\footnote {Note that for $S^n$ and large $n$  the action is more
divergent $S\sim -1/\Lambda^{n/2}$ and it is suggested one should
minimize the energy to help predict $n=4$ [178].} For negative
$\Lambda$ the action is positive and infinite for non-compact
cases [179,180]. One might expect to take the minimum volume of
$V_4$ in this case to minimize the action. This in turn will
restrict the 3 geometry of any created Lorentzian universe  to
also have small volume as in Weeks and Thurston spaces -see [181]
for reviews of hyperbolic geometries in cosmology.

 The wavefunction $\Psi_{HH}$ for de Sitter is therefore
 $\sim \exp(1/\Lambda)$. If $\Lambda$ is given by a distribution function then such expressions would suggest
 $\Lambda\rightarrow 0$: this was at one time part of the wormhole ``big fix'' approach of
 Coleman [182,1]: a wormhole here being a possible connection
  or interaction between different universes. The inclusion of four form field strength
 $F_{\mu\nu\rho\lambda}$ can alter this prediction [183,184]. Recently a decoherence term has been invoked to
 also alter the prediction [185]. However,  this term would correspond to
 adding
 an imaginary part to the scalar potential so causing absorption cf. e.g.[186].
 Since we are presumably dealing with a single universe it is unclear
 how such absorption  can be interpreted, although decoherence has been invoked to explain why different
 possible branches of the universe fail to interfere [187,188]. Back reaction effects from the presence of
 other matter sources  could also affect semi-classical tunnelling predictions cf. [189].   We would also
 add that related earlier investigations using quantum cosmological reasoning suggested a small $\Lambda$ could
 be expected so possibly coinciding with the present universe [190].

 In the tunnelling approach the action is expected to be positive
 definite. This can be achieved in a path integral formulation by imposing ``outgoing only''  modes
  e.g.[169]. In simple models one can
 Wick rotate, the gravitational term in the action, in the opposite
 direction to that used on any matter component. In
 the simple cosmological model the wavefunction is now $\Psi=\exp(-|S|) =
 \exp(-1/\Lambda)$ [163]. This rotation has been justified more
 rigourously from path integral methods, in that the wavefunction should not be
 exponentially large [191]. Although other wormhole solutions could violate
 this requirement [192].

 In analogy with the field emission of electrons problem, where
 the applied electric field $E$ gives the probability of emission
 $\sim \exp(-1/E)$, the tunnelling boundary condition suggests a
 large value of $\Lambda$ is expected.

 We have questioned whether
this analogy of treating the universe like $\alpha$ decay or a
Scanning Tunnelling microscope is entirely sensible [7]. In these
atomic examples the various particles already exist. But now the
``particle'' is the universe itself coming into existence - this
is claimed not to be possible with the Copenhagen interpretation
of quantum mechanics [193] but might be formulated in terms of
generalized quantum mechanics [156]. Note also that the barrier to
be tunnelled through is dependent on having closed $k=1$
curvature. It is absent  for the flat or open cases, which
classically can tend towards zero scale factor in the distant
past. Although such models can have apparently infinite size and
so infinite action one can instead topologically compactify the
space at arbitrary small size [164,194,130]. For such cases one
can impose, for example, analogous outgoing like boundary
conditions but difficulties are present for the Hartle-Hawking
case [194].

{\bf 3.2 Massless scalar field case}

 The previous barrier in the WDW potential is also
dependent on having a strong energy condition violating matter
source. For the alternative example we next
 consider the massless scalar field in a closed
 FRW universe, given by the Wheeler-DeWitt (WDW) equation, e.g.[1-3]. We
 follow our earlier presentation of this example [195],
 \begin{equation}
 \left (\frac{\partial}{\partial a^2}+\frac{p}{a}\frac{\partial}{\partial
 a}-\frac{1}{a^2}\frac{\partial}{\partial \phi^2}-a^2 \right ) \Psi(a,\phi)
 =0
 \end{equation}
 where $p$  again represents part of the factor ordering ambiguity.
\begin{figure}
\begin{center}
\includegraphics[width=6cm] {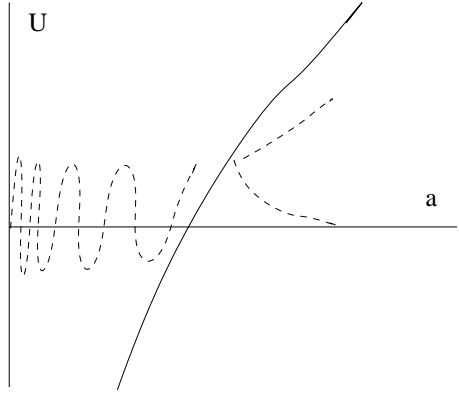} \caption { The WDW
potential (solid line)  for the massless scalar field. Now a
forbidden region occurs, due to curvature $k=1$, at large scale
factors with corresponding exponentially growing and decaying
solutions. The oscillating solutions gain arbitrary short pitch as
$a\rightarrow 0$ representing a singularity with diverging kinetic
energy. }
\end{center}
\end{figure}
The WDW potential now has a forbidden region at large scale
factors,  where the classical  closed universe starts
re-collapsing - see Fig.(2).

 The WDW equation can be separated to
 \begin{equation}
 \left ( \frac{d^2}{da^2}+\frac{p}{a}\frac{d}{da}+\frac{\nu^2}{a^2}-a^2  \right )\Psi(a)=0
 \end{equation}
 \begin{equation}
 \left (\frac{d^2}{d\phi^2}+\nu^2 \right ) \Psi(\phi)=0
 \end{equation}
 with $\nu$ the separation constant.\\
 The solution to these equations can be obtained using e.g. MAPLE [196],
 \begin{equation}
 \Psi(a) \sim a^{(1-p)/2} \left\{ \alpha J_{i\nu/2}(ia^2/2)+
 \beta Y_{i\nu/2}(ia^2/2) \right\}
 \end{equation}
 \begin{equation}
   \Psi(\phi) \sim \exp (i\nu\phi)
   \end{equation}
 where $J$ and $Y$ are Bessel functions (see e.g. [197])
 and each term has an
 associated arbitrary constant $\alpha,\beta$, which we can choose accordingly.

 First consider the limit $a\rightarrow 0$.
 Using the asymptote $J_{\mu}(z)\sim z^{\mu}$ as $z\rightarrow 0$
 enables the solution to be expressed as
 \begin{equation}
 \Psi(a)\sim a^{(1-p)/2} \exp (i\nu\ln a)
 \end{equation}
 There is a divergence as $a\rightarrow 0$ producing
  an infinite oscillation representing the classical singularity
 as the kinetic energy of the scalar field diverges [198]. A similar divergence  also occurs
 for $\phi\rightarrow
 \infty$.

But  as it stands, the solution can  be regularized by integrating
over the arbitrary separation constant. Now the integral
\begin{equation}
\Psi(a,\phi)\equiv \Psi(a)\Psi(\phi)\sim \int  \exp \left
(i\nu[\ln a +\phi] \right )d\nu
\end{equation}
is of the form $\int \exp (ixt)dt$ which by means of the
Riemann-Lebesgue Lemma tends to zero as $x\rightarrow \infty$ (see
eg. ref.[199]). The wavefunction is now damped as $a\rightarrow 0$
or $\phi\rightarrow \infty$. There is another possible divergence
for factor ordering $p>1$ but we have assumed its coordinate
invariant value of unity [200]. It has also been suggested that in
the context of wormhole solutions these milder divergences due to
the factor ordering are not particularly serious [201]. They are
also present for flat empty space, so they conceivably anyway
should be renormalized away [7]. Further considerations of factor
ordering on the various proposals are given in [202]. For large
scale factor the wavefunction eq.(28), behaves as $\sim
\exp(-a^2/2)$ so indicating asymptotically Euclidean space
cf.[198].\footnote{Provided the combination $J+iY$ which equals
the first Hankel function $H^{(1)}$ [197,198] is chosen.}

In general one can impose a boundary condition like De Witt's
original  $\Psi(a=0)=0$ suggestion [67]. As it is  a bound state
problem, square integrable wave functions can be obtained so
giving a normalizable wavefunction cf.[203]. However, this
boundary condition imposed at short distance does not necessarily
determine the behaviour at large distance where possible growing
and decaying exponential solutions are present cf.[204].

For later use we can also mention, as an example of scalar-tensor
gravity, the Brans-Dicke model which
 is derived from the following action e.g.[32]
\begin{equation}
S=\int d^4x\sqrt{g}\left ( \phi R-\frac{\omega}{\phi}(\partial _
{\mu}\phi)^2\right )\;\;.
\end{equation}
For stability in Lorentzian space one requires $\omega>-3/2$.

Using standard techniques, the corresponding WDW equation can be
obtained [195]. Only the equation for
 $\Psi(\phi)$ differs from the previous case
 \begin{equation}
   \Psi(\phi) \sim \exp (i\sqrt{B}\nu\ln \phi)
   \end{equation}
   for $q=1$ and $B=(3+2\omega)$. So the oscillatory divergence now occurs for
   $\phi\rightarrow 0$ as well. Again one can integrate over the separation constant
   to produce a regular wave function

 Since  in the limit $\phi\rightarrow 0$  the Planck length $l_{p}=G^{1/2}\rightarrow
 \infty$, you might have expected this divergence. There is also
 an increasing pitch or ``wiggliness''  of oscillation
 in the
 $\phi\rightarrow \infty$ limit: now at large distance beyond the Planck
 length. Although less severe than the previous divergence
 the pitch can still develop at arbitrarily short distance. We can expect
 these   rapidly oscillatory wavefunctions more
 generally for
 scalar-tensor gravity models, including  non-minimally coupled scalar fields cf.
 [205]. Also higher order correction to the gravitational action
 typically correspond to additional scalar fields in the Einstein
 frame, see e.g.[206].

 One can see a similar behaviour in the WDW solution for a pure
 cosmological constant. The pitch of the solution gets
 increasingly shorter at large distance $a$ cf.(Fig.1). This is simply
 because the universe keeps accelerating and so the ``velocity''
 $\dot{a}\rightarrow \infty$. In general although the
 $\Psi(a)$ part of the solution might be regularized at short distances the actual
 solution $\Psi(a,\phi)$ can oscillate at arbitrary short pitch
 due to the kinetic energy of the matter component. Only simple ``on-shell'' perfect
 fluid models allow the matter to be expressed in terms of the
 scale factor. For the FRW case the scalar field has an extra
 degree of freedom over a perfect fluid e.g.[207]. Note also that both
 signs for the kinetic term are automatically included without
 further restrictions on the separation constants e.g.[208].

{\bf 3.3 Scalar potential $V(\phi)$ case}

We now generalize to a inflationary scalar potential. In the
common boundary conditions, such as the no boundary or tunnelling
ones, the massless component matter is disfavoured or suppressed.
This therefore allow any inflationary matter present to become
dominant during the early stage of the universe. The separation
constants are effectively forced to be zero by the imposition of
the boundary condition which in turn prevents any  kinetic energy
component from causing  an infinitely oscillatory wavefunction.

Approximately, the two boundary conditions give the same results
for $V(\phi)$ as the earlier  cosmological constant case but with
$V(\phi)$ taking the place of the $\Lambda$. On therefore obtains
$\Psi_{HH}\sim \exp(1/V(\phi))$ and $\Psi_T\sim \exp(-1/V(\phi))$
for the Hartle-Hawking and tunnelling cases
respectively.\footnote{ The HH case was extended to complex
metrics by Hartle [209] and another tunnelling type proposal is
given in [210].} One can obtain conditional probabilities for the
initial value of field $\phi$. The tunnelling boundary condition
peaks at large values of $\phi$ so is more inflationary than the
HH case [1-3]. It has been argued that if one allows arbitrary
large energy densities then even the  HH case can also give
sufficient inflation [211] - but one is working beyond the strict
semiclassical domain of validity . Also other matter sources like
the four-form or boundary effects can alter this prediction.
Particle creation effects could also occur during the tunnelling
process [212]. See also [213] for further debate on the relative
merits of the various proposals.

It would seem preferable  to try and work without a particular
boundary condition in the absence of strong reasons for a
particular choice. This is the usual ``principle of indifference''
in the absence of further knowledge. Grischuk and coworkers
[214-216] tried to obtain the average solution for the WDW
equation with a massive scalar field source. They found that the
typical solution was more like the tunnelling one, in that
$\Psi(0)>\Psi(1/\sqrt {V(\phi)})$ so suggesting a tunnelling from
the origin. The HH case corresponds to an exponentially increasing
solution from the origin.  Also with some arbitrariness in putting
a measure on a ``sphere of quantum states '' they obtained a
similar measure at the quantum boundary as previously obtained by
classical equipartition arguments: this strongly favours
inflationary behaviour - see section (2.1) . One might hope that
this sort of argument can be generalized to other matter sources
and when inhomogeneous degrees of freedom are present . Some more
realistic supersymmetric matter sources are explored in [217].

 One
might still  question whether this strong energy matter
suppression is reasonable and indeed it has been suggested that
``zero point'' fluctuations  alone could  alter this picture
[218]. We represent this in Fig.(1) by the solid arrows where
quantum effects will cause the potential $U$ to be ill defined and
allowing a Lorentzian region to also develop around $a=0$. A
classically allowed region near the origin could therefore result.
It is also the case that the presence of strong energy violating
matter is a strict requirement for these common boundary
conditions otherwise no natural Euclidean or forbidden region
would be present. More generally it is suggested that the surface
between Lorentzian and Euclidean regions should have zero
extrinsic curvature $K_{ij}=0$ : so called {\em  real tunnelling
geometries} [219]. In simple geometries this requires being at a
stationary point in the scale factor i.e. $\dot{a}=0$. So the
surface is either the bounce point (cf. de Sitter ) or at the
maximum in a closed re-collapsing model. In ref.[220] the HH
approach was applied to start the universe at its maximum size
before then collapsing.

 Other predictions that distinguish  various interpretations
 of quantum mechanics have been investigated,
  Bohm ``pilot wave'' e.g. [221] or  many-worlds [222]. A further extension to
 field theory  ``universe'' creation operators  or so-called 3rd quantization has
 also been pursued [223], but criticized in ref.[169].

 {\bf 3.4 Further Topological and Geometric aspects}

There is also the issue of including different topologies and
geometries for the tunnelling amplitude in the more general case.
However, proving whether  4-manifolds differ and so knowing if
they are being over-counted in the path integral is undecidable.
It has been suggested that the wavefunction should not be such  a
turing non-computable number [224]. Although whether this would
have serious limitations on using $\Psi$ is disputed since we only
require limited accuracy in our predictions [225]. One can also
work with conifolds: and other more general notion of manifold to
help alleviate this problem [226].

 If the number of
manifolds for the hyperbolic case can approach infinity it can
overwhelm the usual suppression factor for the creation of a
single universe with a given set topology [227]. The ``average''
topology might be able to predict the spatial homogeneity of the
universe [227,228]. But again is this really reasonable? It
implies an initial state or reservoir of all infinite possible
topologies that should be included in the amplitude. There are now
infinitely many ``particles'' one for each possible topology and
geometry. Neither is it clear why just a single universe with an
average topology results and not that many universes each with
different topology form together. Working with closed models, and
so fewer possible topologies, see e.g. [229], could alleviate this
problem but the notion of curvature itself will also become hazy
at the Planck scale.

Indeed the way that curvature  is treated as a constant is rather
unsatisfactory. In FRW models the actually   local characteristic
$k$ is taken to be globally constant. In more general metrics the
curvature can become a function also of time and space $k(t,x)$
cf. Stephani models e.g.[35]. For the  FRW model with perfect
fluid $p=(\gamma-1) \rho$ the WDW potential takes the form,
e.g.[201]
\begin{equation}
U=ka^2-Aa^{4-3\gamma}
\end{equation}
where the constant $A$ can be obtained from the relation
$\rho=A/a^{3\gamma}$. For a forbidden or Euclidean region at small
scale factor $a$  requires $U>0$ which requires $k=1$ and
violation of the strong-energy condition i.e. $0\leq \gamma<2/3$.
However, in a more general inhomogeneous model this behaviour can
be drastically altered. For example in a Stephani model the
corresponding WDW potential becomes cf. [230]
\begin{equation}
U=\beta a^n-Aa^{4-3\gamma}
\end{equation}
so for $n>2$ the forbidden region can be either narrowed or absent
entirely  even for closed models $\beta>0$ and when  the strong
energy condition is being violated. This example is symptomatic of
what, more realistically, can be expected as the Planck epoch is
approached. Non-minimally coupled scalar fields or the presence of
large anisotropy can also remove the barrier [231]. The presence
of forbidden regions that play such a prominent role might then
actually be absent  even for closed models that can display
inflation. Another possible complication is that a negative energy
density can also create a forbidden region.
 For example in a toroidal
model with compactification scale $L$, one typically obtains a
Casimir term, e.g.[232]
\begin{equation}
\rho=<T^0_0>=-\frac{\alpha}{L^4a^4}
\end{equation}
with $a$ the scale factor and the constant $\alpha$ depending on
the nature and number of matter fields present. More elaborate
twisted scalar fields can also be possible [233]. This Casimir
term corresponds to a $\beta>0$ and $n=0$ term in eq. (35) [164].

Even without a forbidden region some boundary conditions might be
adapted to purely Lorentzian metrics, although the underlying
principle is then often less prescriptive cf.[194]- where the
``outgoing only '' aspect of the tunnelling boundary condition was
implemented in such a case.

Further examples of the WDW equation with  variable constants
[234] or variable space dimensions [235] have also been done.

{\bf 3.5 Arrow of Time and classical description}

In order for quantum cosmology to give a realistic description of
the seemingly classical universe a number of conditions have to be
met. The quantum calculation typically gives an ensemble of
solutions [1-3]. One requires that with suitable coarse graining
there is no interference between them; a decoherence functional
$D(h,h')$ has been developed for this purpose [156]. Likewise
expanding and collapsing branches should be independent, which
appears the case for suitably sized de Sitter spaces [187]. How
all the various solutions decohere after quantum  tunnelling is a
further complication cf.[236].

For inflationary models we require quantum fluctuations to source
classical perturbations of amplitude $\sim 10^{-5}$. Initially the
quantum fluctuations are in their ground state for Hartle-Hawking
[237,238] or Tunnelling boundary conditions [239]: so that
standard quantum field theory on a fixed background can be applied
to calculate growth of perturbations [52]. Incidentally, although
there is a gradually growing mode for modes leaving the horizon
[238], it seems too slow to explain why the universe today could
start accelerating cf.[240]. See also ref.[241] for further
considerations  of decoherence of  quantum fluctuations to give
classical ones.

The growth of perturbations can help explain the time-asymmetry of
the universe [242,243]. Typically the boundary conditions are time
symmetric and produce an ensemble of also time symmetric classical
solutions see e.g.[244]. If entropy increase is correlated with
growing scale factor this suggested that a collapsing universe
would have a reversed arrow of time [245]. However, an individual
solution does not necessarily display this symmetry because of
growing perturbations and so a final big crunch can be arbitrarily
disordered: this agrees with the notion that the Weyl tensor is
correspondingly large at a big crunch singularity [24,25].

The boundary conditions can implement a low entropy state and so
produce a thermodynamic arrow of time ( 2nd law of
thermodynamics). Other approaches have stressed that a more time
symmetric approach is warranted and that a final big crunch
singularity should be treated just the same as the initial smooth
state - a time symmetric boundary condition [18,245]. This seems
to questionably  require the semiclassical equations to also break
down at the turning point of a closed universe so that quantum
phenomena then can conspire to decrease the entropy during the
subsequent collapsing phase: quantum wavepackets appear to
disperse at the maximum size [246,16]. This could cause black
holes to re-expand during the ensuing collapse and avoid a loss of
information [18,247]. We later will address another idea for
maintaining unitarity during black hole evaporation.

Using a  density matrix $\rho_i$ for the initial state of the
universe [248] the time asymmetry can be described by postulating
a final indifference principle $\rho_f\propto I$ where $I$ is the
unit matrix ( so summing over all possible final states)
[156,249]. Whereas with a more constrained final boundary
condition one might expect to see more unusual quantum phenomena
occurring perhaps in violation with known results [249,250].

{\bf 3.6 Brane quantum cosmology}

We here just wish to address a number of complications that brane
models appear to have when they are assumed to be created from
nothing by quantum processes. Recall that usual matter is confined
to a brane existing in a bulk space. In the Randall Sundrum II
model the brane is embedded in 5 dimensional Anti-de Sitter (AdS)
space [92].

i)  One can consider either a pre-existing bulk with branes being
created or the bulk and brane appearing simultaneously [251,252].
Some related schemes are given in [253]. A similar situation can
occur with multidimensional models with tightly compactified extra
dimensions where {\em external nothing} and {\em total nothing}
were distinguished [254]. If there is already a pre-existing bulk
it can help fix the relevant outgoing only tunnelling boundary
condition [252]. However, if branes can be spontaneously produced
within a pre-existing bulk there are questions as what production
rate is possible? A similar concern was present in earlier ideas
that the universe spontaneously occurred in Minkowski space
cf.[157,158]. Without some removal mechanism an infinite number of
branes would eventually be produced within a fixed bulk.

ii) If we also require creation of the AdS bulk space then one
needs to use a rather convoluted  ``cut and paste'' procedure to
produce a compact space with a Cauchy surface [96]. Recall the
geodesically complete  closed AdS space is not globally hyperbolic
[28] and boundary conditions would be required to regulate
continuously the time-like infinity.

 iii) For creation of say compact de Sitter branes in Euclidean AdS
 space we have a mixture of negative action (de Sitter) and
 positive action (AdS) parts. Can  these be  simply added or
 should the individual actions be first rotated to positive values
 as expected for tunnelling processes. One could ``cut and paste''
 two manifolds with equal and opposite action so allowing
 unlimited creation possibilities. This suggests that action alone
 might not convey the true complexity of such processes.

 iv) Usually we might expect Euclidean space to be confined
 to small size. But if a large AdS space is also  produced the
 Euclidean nature is being allowed to arbitrary large scale. This
 is related to the previous point about cancelling different
 actions. However because the bulk is  a static space it can obey the
 real tunnelling condition  that the surface between Euclidean
 and Lorentzian has no extrinsic curvature cf.[219].

 v) The brane can have an induced curvature from being embedded
 in the higher dimensional bulk. The  full action can have extra
  terms that, although not contributing to the equations of
 motion of the brane itself, affect the quantum
 calculation cf.[255,256]. It is also important that
 the brane remain  compact to prevent an extra surface term, something
  not automatically resulting from taking a compact bulk space.

A number of extra issues therefore have to be clarified  before
any calculation of brane creation can properly proceed. For this
reason one can instead attempt to promote the branes to an eternal
existence just as the static nature of the bulk space would
suggest. This leads one to cyclic universe models with possible
brane collisions. One example again is the ekpyrotic scenario
[49]. We would just mention that we do not think that entropy
growth has been adequately dealt with in this, and related
scenarios,  and that the  models will generally suffer the same
problem as the earlier Tolman oscillating model. Inflation is used
as an entropy sink, but inflation does not actually reduce entropy
which would contradict the generalized 2nd Law of Thermodynamics
[136]. The problem is closely related to whether information is
destroyed e.g.[257] or preserved by black holes [258]. The problem
is further compounded by using an already infinite sized universe
{\em ab initio} to hide the problem of entropy production. This is
also done in ref.[76] where continual expansion is apparently used
to dilute the entropy in the universe to effectively flat space.
For a finite sized universe continually increasing the entropy
would eventually surpass the critical density and so produce
re-collapse to a future big-crunch singularity. It therefore seems
a sleight of hand to claim that flat space is the maximum entropy
state cf.[76].

 {\bf 3.7 Universe from a quiescent or static state}

We have spoken of the universe starting from nothing or by
bouncing from a previously collapsing phase. A third possibility
is that originally the universe was initially stuck in some
unchanging quiescent state. Perhaps involving the presence of
closed timelike curves (CTCs), see e.g.[30] for a review. This is
closely related to introducing a topological identification scale
as in Misner space e.g. [30].

Starting with Misner space, Gott and Li [218,259] obtained a
self-consistent adapted Rindler vacuum state for a conformally
coupled scalar field that remains finite at the Cauchy horizon,
unlike for the Minkowski case [260]. They then conformally
transformed this state to give a suitable vacuum state for
multiply connected de Sitter space. Such a de Sitter space with
CTCs might then  be an  initial state for the universe. It only
has retarded solutions so giving a possible arrow of time and is a
state of low entropy, actually of zero temperature [218] .

However, in Misner space this state was only possible with
identification scale $b=2\pi$, or $b=2\pi r_0$ for the multiple de
Sitter case [218,259] . Such an exact value is itself inconsistent
with notions of quantum uncertainty. We are therefore wary of
claims that such a multiply connected de Sitter state is stable
especially since the relevant time loop is approximately $\sim$
Planck time, only a plausibility argument has so far been made
[261].

The actual procedure of balancing a negative starting vacuum with
a Hawking radiation, due to the periodicity, to give an empty
vacuum state has further possible difficulties. The calculation
makes use of the periodicity producing a thermal state [262] .
Such a state is required to be a many particle state with
technically a suitably large Fock space, see e.g.[111]. But by
being close to the Planck scale one starts reducing the number of
allowed states due to holography type arguments [77]. This will
start preventing an exact thermal state, as also is expected
during the final stages of black hole evaporation [263] or in
Planck scale  de Sitter space [39,264]. This mismatch could then
result in some fluctuations still being present in the vacuum
instead of a pure empty state, so destabilizing the CTC.

 Neither is
it clear that the $b$ value, or the corresponding de Sitter one,
remain independent of different matter couplings $\xi$ or
potentials $V(\phi)$. A more realistic combination of matter
sources still appears divergent at the Cauchy horizon [265],
although an improved {\em self-consistent} renormalization
procedure [266] in Euclidean space might help regulate some of
these other cases.

 Creating this state in any case seems rather contrived. Recall
that the Rindler vacuum of accelerating observers requires
``mirrors and absorbing stray radiation'', before we then make any
topological identification [267]. One would need some more general
reason why such an initial state was actually present. The
analogous zero temperature state for charged Black holes has
proved difficult to obtain on grounds of stability [268].

Instead of requiring CTCs one might just allow a static state with
time still evolving normally from say $-\infty$.  There is a
recent emergent model [269], an update of the
Eddington-Lema\^{i}tre model e.g. [31] that starts from an
Einstein static universe. Because this model has no forbidden
region, and requires a balance of ordinary matter and a
cosmological constant, it again will  be prevented by the usual
boundary conditions that bias against the normal (e.g. radiation)
matter component. Neither do we think that maximizing the entropy
is a more suitable principle for determining the boundary
condition since the entropy actually grows later during the
inflationary stage cf.[264]. The emergent model is however
geodesically complete to the past unlike the previous case of
multiply connected de Sitter space. One might do a further
identification of antipodal points in the multiply connected space
but one then loses time orientability. This would make the problem
even more involved cf.[270]

Also such a model also requires a mechanism to stabilize the
Einstein static phase to homogeneous perturbations, although the
model appears stable to inhomogeneous perturbations [271]. More
general inhomogeneous models might allow this. For example, by
altering the curvature dependence as in eq.(35) one could produce
a stable static universe with a now flat $U=0$ WDW potential; or
perhaps, at least prevent collapse to the origin by means of a
repelling potential $U>>0$ around the origin $a=0$,  cf. a Casimir
term [164].

 Such a repelling or
Planck potential has been added previously in an {\em ad hoc}
manner to produce a forbidden region around the origin
[272,18,247]: a term $\propto a^{-2}$ is added to the WDW
potential $U$. This can then allow a De Witt like boundary
condition $\Psi(0)=0$ [67]. Also a non-zero separation constant
can display a similar effect [273] cf. eq. (26). Since, as
previously mentioned, a  massless scalar field automatically
includes such separation constants it can alone be used to produce
a bouncing model as in [274]. It is also claimed that a perfect
fluid can give such behaviour [275], but this would require an
asymmetric Wick rotation cf.[208].
\begin{figure}
\begin{center}
\includegraphics[width=6cm]{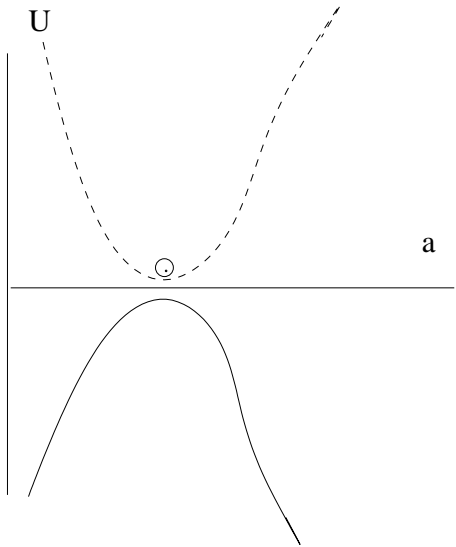}
\caption{ The WDW potential (solid line) for the unstable Einstein
static universe. Reversing the potential (dotted line) would
produce a stable static state.}
\end{center}
\end{figure}

 One might also try to stabilize the Einstein static universe
more generally by surrounding the state entirely with forbidden or
Euclidean regions. For example if the sign of the WDW potential U
is flipped the corresponding Einstein static universe becomes
stabilized - see Fig.(3). Such a model requires $k;\Lambda;\rho;$
$\rightarrow -k;-\Lambda;-\rho$, so now this is an open AdS with
an extra negative radiation component. So violation of the weak
energy condition is now required for such stability, such as might
occur with the Casimir effect.

 A stabilized state could also be achieved
without altering the matter component by use of a signature
change, represented by the parameter $\epsilon$: $\epsilon =1$ for
usual Lorentzian space and $-1$ for Euclidean space [276]. In the
simplest case the corresponding WDW potential is altered
$U\rightarrow \epsilon U$ [207]. So if  for some reason
$\epsilon=-1$ the previous static universe is again  stabilized.
Other possible examples of signature change, starting from
different action principles, are also possible [277].

It has been suggested that oscillations of $\epsilon$ between the
two cases are constantly occurring but that the ``average''  now
favours a  Lorentzian space-time [278]. One might imagine instead
a preponderance of negative Euclidean values for $\epsilon$. This
might help stabilize a static model before for some reason the
sign changed and Lorentzian evolution then could proceed.

Despite these present difficulties the notion of finding  a
suitable quiescent state has some attraction. The difficulty, as
in the examples given, is why the state should survive for
semi-infinite times, but still have some slight instability that
causes the expansionary or Lorentzian evolution to eventually
begin. For example, a previous model [279] suggested eternal
oscillations around the Planck size could exist but it did not
then produce a large universe in the infinite future. Also recall
in section (2.1) a scalar field model that might oscillate a few
times before undergoing inflation.

{\bf 3.8 Black hole final state}

Black hole evaporation seemingly produces a  mixed from  a pure
quantum state [263,280]. This produces an increase in the von
Neumann entropy, so agreeing with a generalized 2nd law of
thermodynamics, the increase in entropy corresponds to a loss of
information beyond the black hole horizon e.g.[111]. When a black
hole evaporates does this information remain lost or does it
perhaps get encoded on the Hawking radiation? A number of possible
ideas have been suggested - see e.g. [281].

One recent proposal that maintains information (unitarity) during
black hole evaporation  is to propose a final state boundary
condition at the singularity [282]. Previously a final
indifference principle was used at the singularity to show that
the Hawking radiation was thermal [280]. Instead a maximally
entangled final state is proposed: this effectively has no
disorder and zero entropy. A process akin to quantum teleportation
can then enable information to be carried by the outgoing Hawking
flux as infalling states are ``measured'' at the singularity
[282]. We just wish to mention some concerns with this proposal
that mostly were previously known for quantum cosmology.

a) One first might question that entangled states are rather
fragile, see e.g.[283], and near a large mass would be expected to
decohere cf.[24]. They also appear susceptible to acceleration
[284]. In ref,[285] it was pointed out that interactions between
infalling Hawking radiation and infalling matter  could prevent a
maximally entangled state and some information would remain lost.
However, the entanglement of the final state was found unnecessary
[285] provided a sufficiently random interaction $U$ between the
infalling matter and radiation occurs. Within a suitable measure a
random state is almost perfectly entangled.

However, the dimension $N$ of the Hilbert space ($N=\exp(S)\sim
\exp(60) $ with $S$ the entropy of the black hole [263] )is so
large that the number of entangled components or interactions $U$
vastly exceeds the allowed limits on total information content of
the universe using holography arguments : $10^{120}$ operations on
$10^{90}$ bits [287,69]. It therefore seems rather unrealistic
that such elaborate states are actually present.

b) In quantum cosmology final state proposals can cause quantum
phenomena to be constrained; so quantum randomness is rather
limited [249,250]. Algorithmic complexity is transferred into the
initial and final states. We have argued that this vast complexity
now transcends that expected for the entire universe. Although the
black hole final state is not affecting the arrow of time globally
it, does appear to allow unwanted phenomena like faster than light
signalling [288]. It therefore seems consistent with earlier
worries about imposing final state boundary conditions in quantum
cosmology.

c) A final concern is that the proposal removes any objective
notion of entropy for the black hole e.g. [111]. Black hole
collapse and evaporation does not increase the entropy ( pure
state $\rightarrow$ pure state ) and entropy is once more a rather
subjective notion of what information is available to arbitrary
observers. One might also wonder how  the analogy  between  de
Sitter entropy and black hole entropy is affected.

 {\bf 4.0  Loop Quantum Cosmology}

Another formulism for quantum gravity is the loop  approach that
treats gravity in a similar way to other gauge theories. One works
with holonomies and Wilson loops, previously used in
non-gravitational field theories [14]. A number of reviews have
been done on technical issues involving this approach [14,15,22].
Earlier work obtained exact solutions of the WDW equation using
Knot theory invariants: the so-called Kodama state [289]. So far
these models have proved difficult to interpret or implement in
cosmological model beyond a simple cosmological constant [290] or
slow roll scalar potential case [291]. A number of related
approaches to quantum gravity have emphasized the discrete nature
of space as you approach the Planck scale [15,22]. This
discreteness might regulate singularities usually signalled by
infinite oscillations in the wavefunctions . This might also alter
the actual application of quantum mechanics, for example an
alternative Bohr quantization scheme was claimed to also give a
more effective avoidance of singularities [292].

We will mostly consider the Thiemann approach [14], within loop
quantum gravity, that has most  been developed  to a level that
can make contact with realistic cosmological models [13].
 One claimed
 advantage of loop quantum gravity over strings is that a
 background independent formulism might easier  be achieved [14]. However
 this might actually be a hindrance for
 realistic cosmologies cf.[293] where
  {\em restricted covariance} was considered more realistic.  Another possible
 drawback  is that in GUT theories the various forces of
 nature should eventually unify. Therefore the present weak force of gravity
 should increase with energy scale to eventually coincide with
  the other forces of
 nature. The Planck length $\sim G^{1/2}$ will correspondingly grow as
 the unification scale is approached. But in loop gravity this aspect of
 ``running'' Planck length does not appear   incorporated at
 present. A
 different initial Planck scale would  correspondingly  alter the
 various cosmological {\em puzzles} and so possibly alter the
 amount
 of inflation required.

 At first sight any  possible discreetness around Planck scales  does not appear
 too helpful in actual
cosmology  since, in particular, flat or open FRW models are
infinite from the start, so the granularity is never actually
apparent. One can make a cut to enclose the big bang within a
finite volume as time $t\rightarrow 0$, but then the matter or
energy momentum becomes infinite within such a domain [294] . By
making a compactification, at a scale $L$, this infinity can be
alleviated to some extent.\footnote{This scale $L$ seems a further
arbitrary constant cf.[291].} But still the matter densities will
diverge as the universe is evolved back towards the initial
singularity: likewise for closed FRW models. Incidentally
compactifying in this way introduces possible vacuum polarization
and Casimir like effects e.g.[232]. Typically the energy density
$\rho$ would become negative so violating the weak energy
condition, unless supersymmetry intervened to suppress such terms.

{\bf 4.1 Bouncing and  Inflationary model}

 Interestingly, loop quantum gravity is also said to alter, possibly every
matter component such that the weak energy condition is
effectively violated at short distances when the granularity of
space becomes significant. Recall that usually matter is diluted
for an expanding universe or remains constant for an exact de
Sitter solution. Likewise for the earlier Casimir  component
eq.(36).

 By allowing the energy density to grow during expansion  there
need not be a divergence in the energy density as the universe is
evolved back in time. One possible advantage is that now  a bounce
from an earlier collapsing phase might be implemented close to the
Planck length scale. Usually one needs to bounce before the Planck
energy density is surpassed by ordinary matter.

However, if the previous phase of the universe already contained
large amounts of matter and entropy it would not necessarily
require inflation. But then the bounce would have to proceed long
before Planck size lengths are reached in order for Planck energy
densities not to be vastly exceeded. Alternatively if the previous
phase was devoid of matter then it can collapse to the Planck
size. Then a later inflationary phase could produce sufficient
matter and entropy. The previous collapsing phase of the universe
is effectively empty: a similar stage occurs in the pre-big bang
model [104].  Another less likely scenario is that a deflationary
phase (rapid collapse) could remove matter and radiation: but this
would seem to violate standard notions of quantum unitarity
together with the generalized  2nd Law of thermodynamics
e.g.[29,136]. In this case all matter component energy densities
scale as $\rho\propto a^n$ with $n>1$ as the bounce approaches,
and a rapid collapsing or deflationary behaviour would be
possible.

It is  therefore  not entirely clear whether  an inflationary
phase, following a previous collapsing epoch of the universe,
should be implemented. Many of the so-called {\em puzzles} will be
different to that of the usual big bang model and need a more
careful consideration. Recall that inflation can take a ``small
bang'' universe to that of a big bang: so effectively behaving as
a universe amplifier. So only if there is a bounce away from  a
``small crunch'' does inflation have to intervene to give a later
``big'' universe. But then there is a mismatch with the earlier
preceding universe that it was  not collapsing towards a big
crunch before evading the singularity. This can be contrasted with
the general point that producing inflation requires low entropy
conditions [24,25]. Whereas a collapsing universe would be
expected to have increasingly growing entropy for the usual
direction for the arrow of time.

\begin{figure}
\begin{center}
\includegraphics{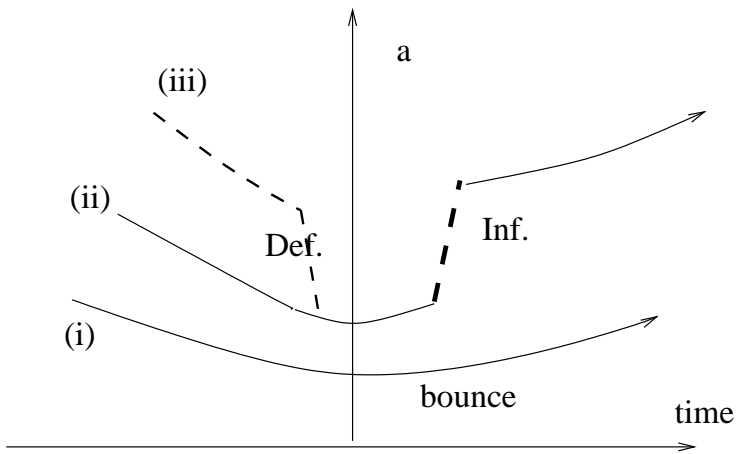}
\caption{ (i) Bouncing universe - matter content symmetrical
before and after bounce; (ii) Bouncing universe with extra
inflationary expansion - presently favoured in loop quantum
approach; (iii) Bouncing universe with deflationary phase and
inflationary phase - allows universe to remove matter and so
bounce at smaller size without violating Planck energy densities.}
\end{center}
\end{figure}

{\bf 4.2 Modified Wheeler - De Witt equation}

We can now consider a few specific issues for the loop quantum
cosmology approach. Again we stick to  the cosmological issues as
much as possible. The scale factor is defined in terms of the
triad component $p$ where $a^2=|p|$ [13,14]. Incidentally, a
similar way of canonically transforming, say the scale factor $a$,
to a new variable, now defined over the full interval $(-\infty
\rightarrow +\infty)$ has previously been used to help regulate or
``cross'' the singularity at $a=0$[161].

At large distance the usual WDW equation should be produced. But
this continuum limit is given by, for a closed model and ignoring
the matter Hamiltonian, [295-300]
\begin{equation}
\left ( \frac{d^2}{d p^2 }-1 \right ) \sqrt{|p|} \Psi (p)=0
\end{equation}
Writing this in terms of $a$ gives
\begin{equation}
\left ( \frac{d^2}{da^2}+\frac {1}{a}\frac{d}{da} -\frac{1}{a^2}
-a^2 \right ) \Psi(a)= 0
\end{equation}

But this now automatically includes a repulsive Planck potential
around the origin cf. eq.(26). It is unclear whether this operator
ordering is actually imposed by the formulism or whether an
arbitrary choice has been made cf.[297]. Partly this is imposed by
``pre-classicality'' that the pitch should not fall below the
Planck length. However a constant inflationary solution develops
arbitrarily short pitch and if  such solutions are prevented it
could, for example,  prevent
 eternal inflation to the future.
  It is claimed [297] that this de Sitter example is only an  ``infrared problem''
  and can be ignored since the local curvature is still small.
   But this distinction
 seems arbitrary and so using the pitch alone would not impose  an
 unambiguous choice of factor ordering.

  If the factor ordering that gives an effective Planck potential is totally justified it can  be
used to produce a $\Psi(0)=0$ type boundary condition [300]. By
introducing a forbidden region around the origin  it also can be
responsible for causing a collapsing universe to rebound cf.
[273]. This Planck potential suppresses any decaying mode
corresponding to $D>1$ in section (3.3) so disallowing any
tunnelling possibility from the origin. Only an exponentially
increasing mode is present in the forbidden region, so the imposed
boundary condition is close to the Hartle-Hawking one cf.[300].

With this reasoning  loop quantum cosmology has given a
justification for earlier ideas about modifying the WDW potential
at small $a$. Incidentally, such a correction to the WDW potential
could alone prevent a Mixmaster phase in Bianchi IX models: since
it is known that the presence of an extra  stiff fluid prevents
chaos cf.[301,302]. Equation (38) corresponds to a stiff fluid but
with the wrong sign. Chaos is also removed by reducing the number
of degrees of freedom, if matter $\phi$ simply becomes a function
of $a$, chaos cannot occur anyway e.g. [303].

At still shorter distances the loop effects cause the WDW equation
to be replaced by a difference equation. One can apparently
iterate to regions with negative p, a so-called orientation change
during the previous collapsing universe phase.
 However, if a forbidden region occurs at large scale as in a
 closed universe model the boundary conditions have to impose the
 relevant solution - see Fig.(2) where the growing and decaying
 modes are sketched.  In order to obtain just the decaying mode one
 needed to chose a suitable Hankel function for the usual massless
 scalar field example. The presence of possible growing
 and decaying solutions at the turning point of a closed universe
 was  previously  found to possibly cause problems with maintaining
 semiclassical behaviour with wavepacket solutions [246].

 For the loop case,
 in ref.[304] they have fixed this solution
 correctly at the maximum size of the closed universe and iterated
 back to the corresponding point in the oppositely  oriented universe. The
 solution there no longer just decreases into the forbidden region. This
 might have been anticipated since the Hankel function corresponds
 to a wave moving in, say, a right direction, so one would need it to
 switch to a left moving direction in the oriented epoch.
 In general it looks at though alterations at short distances does
 not resolve all the ambiguities of boundary conditions. Indeed it
 has previously been emphasized  that quantum behaviour is not
 necessarily confined to small scale factors only cf. e.g.[305,279,245].

{\bf 4.3 Loop cosmology and super-inflation}

 At intermediate scales there is a possible semiclassical effect
 that alters the behaviour of a scalar field matter source. This
 effect occurs up to distances $a_*$ that depends on an arbitrary
 parameter $j$.
 Provided an inflationary phase is still necessary  a new mechanism
 might now be possible.
 The idea is that a massless scalar field can  itself become inflationary. The
 kinetic energy increasing as the universe expands, typically $\dot{\phi}
\propto a^{12}$; this now  corresponds to an effective $\gamma\sim
-3$ so strongly violating the weak energy condition. This produces
pole-law like inflation but alone is  not sufficient to solve the
usual puzzles of standard cosmology: for one thing the final value
of the scale factor $a_f$ should be $\sim cm$: so that the
subsequent universe be  large and produce mass density $\sim
10^{-30} gcm^{-3}$ today [51,55]. This would require the parameter
$j$ [13] to be extremely large so that loop effects are prevalent
up to $cm$ distances. Another problem is that pole-law inflation
has a corresponding growing Hubble parameter that produces a blue
spectrum of perturbations [306,55]. Because the size of
gravitational waves is given by the Hubble parameter it should not
be allowed to become larger than $\sim 10^{-5} m_{pl}$ [307].
Incidentally it might be possible, with quantum gravity effects,
to suppress the fluctuations by a factor $\sim 10^5$ so that one
could work with inflation at the Planck energy scale [308].

It was then suggested that since  the pole-law inflation was not
alone sufficient, a second  potential driven inflation could also
be present [298,300,309]. This is because  the friction term in
the Klein-Gordon equation can change sign and possibly drive  a
scalar field up a potential. However, one can show   that  growth
in $\phi$ is negligibly small while the kinetic term is growing at
its maximum rate. Enormous changes in $\dot{\phi}$ are occurring
for only small eventual values of $\phi$. In order to prevent
kinetic energies beyond the Planck scale being produced, with
allowance made for this large growth in $\dot{\phi}$,  puts severe
constraints on the initial values of $\dot{\phi}$. Typically for
the chosen ($j\sim 100$) parameter the initial value is
constrained such that  $\dot{\phi_i}^2 \leq 10^{-12}$. This value
alone is too small to drive the field significantly up the
potential and so produce sufficient overall inflation. Later [310]
other quantization schemes, with a second arbitrary parameter $l$,
were suggested that can produce other functions of the scale
factor that might also go in front of the potential term
$V(\phi)$. But even then, contrary to the impression of ref.
[310], very little phase space for sufficient inflation is
actually present if the field is not initially displaced from its
minimum. Also other, possibly massless scalar fields, would have
to be suppressed and spatial gradient terms not allowed to cause
instabilities; a first step in showing that inhomogeneity does not
propagate with super-luminal speed is given in [311]. Growth in
$\phi$ is at the expense of suppressing growth in $\dot{\phi}$
[312], so massless scalar fields will tend to be first to reach
the Planck boundary. Incidentally, in Euclidean space the friction
term is also switched in the corresponding Klein-Gordon equation.
This ``anti-friction'' mechanism has been used heuristically in
conventional quantum cosmology to explain  a large initial field
cf. [1].

The mechanism therefore has  negligible effect on the overall
measure for inflation. Recall earlier argument that equipartition
for initial $\rho\sim 1 $, or in term of the Planck mass $\rho
\sim M_{pl}^4$, only failed to give inflation when
$\dot{\phi}^2\sim 1$. But these large values of kinetic energy are
not suitable to improve further the probability of inflation $f_i$
already strongly favoured around $\sim (1-10^{-5})$. However, the
main uncertainty of quantum cosmology is to determine the initial
value of $\rho$. Perhaps for small initial values of $\rho$ the
super-inflation can alleviate the decline in $f_i$ as $\rho$ is
reduced. Working with the purely classical canonical measure does
not resolve  this question since the conjugate variable $p_{\phi}$
just changes its $a$ dependence and a related measure is
reproduced cf. eq.(8): an infinity of solution would surpass  the
Planck energy density and these must somehow be first excluded. An
infinity of under-inflationary solutions is also present. Unless
one can argue that ``repeated tries' are allowed, which is closely
related to an anthropic argument, or else suggest quantum gravity
processes would intervene once the Planck energy density epoch is
again reached. But the rough uncertainty principle in energy/time,
which is here marginally more correct than a non-relativistic
position/momentum uncertainty principle [309,310], already biases
strongly towards inflation: at least for simple homogeneous scalar
field models.

The Hartle-Hawking proposal seems to give a small $\rho$ but the
initial size of the universe is correspondingly large $a_i\sim
\rho^{-1/2}$: this is vastly larger than $a_*$ so it would never
be in the loop quantum regime. But if the model is actually
bouncing not because of a small effective cosmological constant
but rather because of a changing behaviour of the kinetic term
this bounce could occur for size less than $a_*$ . The size
depends on the coefficient $A$ in equation (3) and can be  set $
A\sim 1$ by scaling into the definition of $a_*$ using the Planck
length and Barbero-Immirzi parameter [14]. Incidentally looking at
equation (3) we can see clearly why a bounce requiring $m>n$
proceeds in the closed ($k=1$) case, since now  $ m=2> n\sim -12
$; or by using a negative cosmological constant ($m=0$) in place
of the curvature  cf.[313-315]. For similar reasons one could
start with an emergent model that commences oscillating before
eventually undergoing inflationary expansion [316]. However,
either a) any ordinary matter must be absent: which seemingly
violates expected entropy increase, or b) all types of matter are
altered to grow with a modified behaviour, as say $\rho \propto
a^r$ with $r>1$ during the loop gravity phase.

Normally the kinetic energy is the stiffest matter and
$\rho\propto a^{-6}$ changes its $a$ dependence by a factor
$a^{15}$ so that  $r=9$ [13,309].  It now requires all other
matter sources to have a smaller effective value  for $r$ less
than 9, if such a mechanism is to be an attractor; or to prevent
another source growing earlier to Planck densities before the
kinetic energy itself can grow significantly. Dust and radiation
matter sources appear to violate all the energy conditions
although less drastically than a kinetic term [317]. If spatial
gradient terms are not altered they might be an impediment to
inflation or bounces occurring. This dilemma awaits the
consideration of going beyond simple scalar field examples. One
might also worry that purely gravitational vacuum terms like
gravitational waves, or the Bianchi anisotropy parameters
$\beta_{-}$ , $\beta_{+}$, could alter their behaviour during
possible Kasner evolution. Although it is shown [301] that chaotic
Mixmaster behaviour is prevented by means of the scalar curvature
being bounded, the shear term $\sigma^2\propto a^{-6}$ itself
might be altered. Especially since if the universe is contracting
in one direction (while still expanding in the  two others) this
direction could be affected more by loop effects. Although still
the total volume of the universe remains decreasing.

 In summary, the anti-friction effect only has a negligible
 effect on the overall  likelihood for inflation. If instead, loop effects could
 convert large initial kinetic energies into potential energy, it
 could make a much more significant contribution to improving the
equipartition
 measure for the chances of inflation being present.

{\bf 4.4 Is Loop quantum cosmology unstable?}

There seem a number of dangers in promoting the scale factor into
a true fundamental distance. For a massless scalar field the
energy density $\rho$ is now modified at short distance such that
$\rho\propto a^n $ The energy density now disappears as
$a\rightarrow 0$ and so is almost indistinguishable from flat
space. But this suggests a danger that actually any Planck sized
region is now potentially unstable to this inflationary expansion.

One might try and reason that for a Planck length region, within a
pre-existing universe, to inflate it requires a negative pressure
that will be quickly equalized by the greater average pressure of
the universe outside. This was one of the reasons that creating a
universe in the lab is difficult because of the surrounding
background metric [148]. But while such an equalization is taking
place there is the possibility of a quantum tunnelling occurring
to a new baby universe. This does not supplant the original
universe but disconnects forming a new universe. In standard
potential driven inflation such a scheme requires one to produce a
high energy density false vacuum state that then has a minuscule
chance of tunnelling to produce a new universe [148-150]. But now
any Planck size region automatically could make such a transition
providing topology changes are not forbidden on other grounds, see
e.g.[30] for introduction to topology issues. In the lab it
required huge effort to violate the strong energy condition, but
now if the weak energy condition is continually being violated at
short distances it seems easier  to conceive of such tunnelling.
One might try and quantitatively calculate this enhancement but
there is another ambiguity: there is an arbitrary compactification
scale for flat or open cosmological models. Usually, these cases
have infinite action due to infinite size and are discounted, see
e.g. [159]. But now with a finite volume $V$ and energy density
decreasing with size such universes are not apparently suppressed
on action principles alone, $S=V\int a^n dt \rightarrow 0$ as
$a\rightarrow 0$. In the closed case a forbidden region is present
so that the created universe must start with at least a certain
size cf.[300]. Placing this value beyond the weak energy violating
region might help suppress the universe creation effect but this
would introduce fine tuning.

We can also consider the creation of the original universe {\em ex
nihilo}. Now even in standard quantum cosmology it is not entirely
clear why universes are still not being  created around us. You
can try and argue that the forbidden region creates a barrier that
to observers within the existing universe suppresses further
universe creation e.g.[318]. But this barrier is either absent or
reduced in loop quantum cosmology and also  in some standard
cosmological models. One can further distinguish between creating
an isolated universe and one being formed within a pre-existing
universe  which requires gradient energy around the linkage [319].
If only new unconnected universes are easily created then you can
argue that this is not a significant problem; although it might
seem profligate if the mechanism causes constant, if unseen, new
universe production.

There is second type of potential instability if the universe does
starts inflating. Because we have particle or event horizons we
only have causal contact up to finite distance - see e.g.[28].
During a high energy inflationary expansion this size could easily
be below $a_*$. In other words, only from the perspective of being
outside the universe can  the  scale factor really be defined. To
stop the inflationary expansion within a  Hubble volume the scale
factor needs to play another ``non-local''
 messenger role.  We seem  in need of
 a sort of generalized Mach's principle e.g.[31,34], telling the individual ``granules of space'' how
 big the total  universe has become. A related concern [153] seems present
  in certain variable constant theories, that also use the scale
  factor to determine, for example,  the actual  speed of light value.
  Maybe this adapted Mach's principle only needs to work up to some suitable
  size but it still occurs over distances not normally believed
  causally connected.

 The idea of space being made of discrete quanta can introduce
 further conceptual problems. In an expanding model new cells have
 to be produced to fill in the gaps. But if we make analogy with
 cell division in living organisms, how are cells produced without
 error? Because presumably there is no analogy with DNA, there
 seems the need of providing ``scaffolding'' to force cells to have
 their correct form.  Normally expanding space is simply stretched like an elastic band by the
 scale factor. Even then gravitationally bound systems (e.g.
 galaxies) can drop out of the  expansionary global behaviour
 of the universe: so the scale factor never plays a universal
 messenger role to individual atoms.
 It seems difficult  that  loop quantum cosmology can
 therefore distinguish that normal matter should display weak
 energy condition violation only when the universe is small and not just at
 any small distance.

 Bojowald has suggested to me that, in the above language, there
 now exists a scaffolding preventing such exotic behaviour due to the
 universe now obeying  the ``average'' classical description. We
 are suspicious about how this is propagated and that the
 super-inflationary phase once started need never stop for
 $H^{-1}\leq a_*$. More recently there is indeed work suggesting that sufficiently small black
 holes might be prevented from forming due to such loop bouncing  effects [320].
 One might address  why they do not in turn go on to drive new
 inflationary universes?

 We can also question
 how Einstein's equations $G_{\mu\nu}\Leftrightarrow T_{\mu\nu}$
 have been
 used to interchange  between geometry and matter i.e.
 $a\leftrightarrow \phi$. It appears a rather  ``on-shell'' restriction
  that matter can be expressed in terms of the scale factor
  cf.[207]. For one thing it reduces the number of independent
  degrees of freedom in the problem. Also,
   in more general scalar-tensor gravity, or with higher order corrections
  to the gravitational action, this
   distinction between the geometry and matter
   is even more involved. The total solution can have arbitrary
   high frequency oscillations that cannot easily be
    confined or excluded by discreetness in the scale factor alone.

 Other standard models such as with a simple cosmological constant
 also get arbitrary short oscillation lengths corresponding to
 increasing kinetic energy. We have therefore suggested that this
 property of quantum gravity becoming important  at short-distance is
 not sufficiently universal to resolve many problems. Requiring that also the energy-density be
 approaching large, or even Planck values, might be a more
 suitably and locally defined quantity. Future experiments might
 see whether a discrete structure does exist using gravitational
 waves, gamma-ray spectrometry or tabletop experiments e.g. [15,321-323].

{\bf 4.5 Summary: loop and quantum cosmology}

In conventional approaches to quantum cosmology the matter terms
have to be introduced in a rather arbitrary manner. Although this
would  eventually have to be justified from the ultimate particle
physics action. The various boundary conditions then decide if an
inflationary matter component can dominate during the initial
stages of the universe. Further restrictions on the initial degree
of homogeneity etc. are required since inflation alone does not
fully explain  the subsequent close agreement with the
cosmological principle. Most work has not yet dealt adequately
with models that are large departures from idealized  FRW or
Bianchi anisotropic cases cf.[1,35,324].

Loop quantum cosmology suggests that due to discreetness effects
at short distance any or most matter sources become
super-inflationary, so effectively violating  all the energy
conditions: weak, dominant and strong. This is now  a stronger
prediction of inflation than previously, being somewhat immune to
the matter components present. But, potentially this could have
drastic, or unwanted, consequences although negative energy
densities have not so far been obtained cf.[325]. However, even
such a super-inflationary phase is still not sufficient to remove
any arbitrary initial conditions ( even if imposed at time
$t=-\infty$ before then bouncing at $t=0$) so other principles are
still required to limit the cosmological model: small initial
inhomogeneity etc. So far  mostly FRW type models have been done.
Also the boundary conditions typically  have to  play a role at
large distances not just close to the  Planck scale where
discreteness can be expected to be important. Explaining the
required arrow of time is also dependent on some chosen
non-equilibrium starting point for the model since no underlying
time-asymmetry has so far been found [24]. There are also issues
of choosing a suitable vacuum state cf.[326].

 {\bf 5.0 Conclusions}

 A number of cosmological schemes have been considered that
 typically start from a region where quantum gravity effects become
 dominant. If the action becomes small then one might reason that
 quantum behaviour might be apparent. The corresponding universe
 is typically of Planck size and mass and so requires an
 inflationary phase in order to amplify the universe to a realistic size.

But such ideas are found to have certain {\em fragility} problems:
like the forbidden region being  strongly dependent on how the
curvature behaves, or CTCs requiring  extreme fine tuning.
Analogies with atomic physics, such as  tunnelling phenomena, are
extrapolated to the universe as a whole. Usually the boundary
conditions have been developed apparently with the sole aim of
starting the universe in an inflationary state. One then at least
must include matter sources in the starting mix  that could
produce inflation. Although, there is still some dispute whether
boundary conditions can totally promote the inflationary
component, it would help prevent ambiguities in purely classical
measures for the probability of inflation. However, what preceded
this inflationary state, and why and how it previously evolved, is
far from clear. At present in string theory,  it is even rather
difficult to produce  violation of the strong energy condition,
see e.g. [327], partly because of complications with having extra
dimensions.

Loop quantum gravity provides a different approach in that any
matter, because of discreteness, is promoted to be inflationary
while the universe has a  ``small'' scale factor. This goes beyond
the notion that discreteness due to quantum gravity might have
been expected to help regulate singularities. The loop effects are
effectively violating all the standard energy conditions so
singularities do not appear anyway.  Although work has been so far
confined mainly to using scalar field sources. Potentially this is
a more robust, or definite prediction of inflation, in that
possibly any matter source could initially be chosen. However,
because distance is ill-defined such a super-inflationary phase is
likely to suffer from a number of other problems: one is that
inflation once started would never end if causal distance is
fundamentally more valid than the scale factor {\em per se}.

If the laws of physics preceded the universe  we might have some
chance of describing the origin of the universe cf.[328]. But if
there is only a single universe then we also have fundamental
problems in applying the notions of probability e.g.[329].
Sometimes a pre-existing meta-universe is introduced but its
origin and laws are not explained further. Fluctuations within it,
however remote, are taken as possibly allowing baby universes to
branch off. These might have suitable low entropy so allowing
dynamical evolution in line with the 2nd law of thermodynamics to
then proceed. Whether the meta-universe itself has dynamical
evolution is a further complication cf.[75,76].

 We can finally consider
 some general principles that might be important for quantum
 gravity explanations of the universe.

 $\bullet$ {\em Action Principle.}

 If on evolving back the current universe to a state when  the total action
  becomes small one can expect quantum effects to dominate. One complication is
  that the gravitational action, once suitably regulated, is not positive definite. One then might conceive
  of quantum tunnelling to this early  minimal action regime. However,  this assumes that
   notions gleaned from standard quantum mechanics can readily be extended to the universe as a whole.
   It also requires that a pre-existing state of  nothing still obey the usual laws of physics:
   why such a state exists and whether it can spawn infinite other
   universes  is also to be understood

   In some
 brane type cosmologies, with infinite  initial bulk
space,  the  branes always have large action {\em ab initio}: they
therefore seem less favourable  to   quantum type creation. But if
no other structures with less action are actually allowed, perhaps
on other grounds, they can also be produced provided that quantum
mechanics is not invalidated at large scales: perhaps by emergent
properties cf.[69] . Since there is no notion of ``lifetime''
outside of the universe, arguments about likely and remote have
less validity.

Quantum mechanics can also regulate possible singularities within
cosmological models. But it still requires the  various matter
potentials to be constrained. Alternative quantization schemes
like Bohr quantization could help further alleviate possible
singularities although this is unlikely to prevent all such
examples cf. hydrogen atom in higher dimensions. Large kinetic
energies, corresponding to the pitch or ``wiggliness'' of the
wavefunction, might be prevented because of possible discreteness
of space. However, such rapid oscillatory behaviour is not readily
 confined to short distance in the scale factor {\em per se}. Care is also required if
 classical field equations are
 first used to simplify models, that  although  classically equivalent, can then  alter the deduced quantum
  theory cf.[207].

  The probabilistic nature of quantum arguments seems a mixed
  blessing: it might justify unusual events or  prevent the
  imposition of a single determined history to the universe.
  Sometimes it is  argued that determinism might still be
  present underlying quantum mechanics e.g.[330], or that our universe is again embedded within  some
  vast or superior meta-universe. If quantum mechanics just essentially
  conserves energy but allows a  Heisenberg like ``borrowing before
  payback'' it does not entirely help address why  something is actually  present. Energy
  is anyway ill defined in gravitational systems but in closed
  universes it can be formally defined to be  zero: positive matter energy cancelled by negative gravitational
  potential energy. But this still does not explain why the
  universe should have the {\em potentiality} to occur,
   even if the universe has no overall apparent cost in
    energy to produce . This was referred to as the {\em irremovable problem} in
  ref.[329].

$\bullet$ {\em Entropy Arguments.}

One of the requirements of a cosmological scheme is to allow the
generalized 2nd law of thermodynamics to occur. One might have
expected the initial entropy to be maximized, or perhaps,  at
least the meta-universe to be in a state of high or maximum
entropy. Although the notion of entropy itself is rather
subjective in that it depends on the resolution or coarse-
graining used: generalized  notions of entropy
 are being developed that include algorithmic information e.g.[331].

One difficulty is that the laws of physics appear invariant under
time reversal, although individual solutions do not necessarily
have this property. For example, inflation is an attractor during
expansion but the reverse deflation is an unstable repelling
solution during the time-reversed contraction.  Some have
suggested that time asymmetry  built into the laws themselves is
required to explain the time-asymmetry of the actual universe e.g.
[20,24]. Without this one needs to understand why the entropy
within the observable universe  is far below being maximized
[24,25,90]. Alternatively you can simply point out that, for say a
FRW universe, the entropy has to take the value it does [332] and
this just comes from the various equation. Increasing the amount
of entropy by adding radiation would simply increase the density
parameter above unity i.e. $\Omega>1$.  It then just becomes then
a question of why this model universe and not some other.

Quantum gravity could only help if entropy notions can be extended
to general gravitational models and perhaps that the cosmological
principle or simply inflationary expansion had some obvious
preferred status.

 The notion that a bounce might occur close to the Planck epoch
 also seems to have unforseen consequences. There is the danger
  that if the entropy is correspondingly small  at the bounce
  point then the arrow of time has run backwards during the
  collapsing phase. Or else the entropy must initially be set
  small at the start, presumably  $t\rightarrow -\infty$. One then
  simply transfers the problem of initial conditions to this
  earlier state cf.[24].  Neither is a bounce at small size, violating the holography bound,
   apparently  consistent with quantum unitarity: that
 information should not readily be destroyed.  Not only actual
 matter but the vacuum state itself must also be adjusted. Otherwise,
 vacuum polarization effects (like Casimir and Conformal anomaly)
 would be expected to dominate as the universe became small cf.[333].

$\bullet$ {\em Theory of Everything.}

 Perhaps any final quantum
gravity theory will be more prescriptive as to what values of
various parameters are allowed. On recent example is that the
topology might be restricted to only a few possible cases from
consistency of string theory [334]. On the other hand, there
presently seems an abundance of possible string vacuum states: the
so-called anthropic landscape [335]. But eventually all arbitrary
parameters, for example those that produce microwave background
fluctuations, might be determined {\em a priori} from first
principles leaving nothing to choice, see e.g [336] for this point
of view. At present many variables are measured or imposed {\em a
posteriori}
 i.e. after the fact, while others are given by  distribution functions.

  In such a closely regulated universe it would appear that
 randomness would play a rather subsidiary role.
  The final theory's laws might also be expected to
 display a Ockham's principle like razor conciseness, perhaps having
 incorporating  notions like algorithmic complexity
 or Fisher information for their formulation [337].
  These and related criteria  might eventually play a role in explaining
 the universe's creation cf.[338].

  However, the theory of everything
 might be limited by G$\ddot{o}$del like incompleteness: our
 theories are developed within the universe and cannot take a
 vantage point  independent of  or preceding the
 universe [339]. Incidentally, if the universe evolved from a Planck sized
 nugget then the laws, vis-$\grave{a}$-vis the holography principle,
 cannot even be represented within the universe at that time - this might be
 construed as a {\em reductio ad absurdum} for the principle. It
 also appears difficult to explain the resulting complexity of the
 late universe from such a reductionist approach that starts the
 universe from a simple state with few degrees of freedom cf.[69].

  The incompleteness problem might be alleviated if, for example,  the laws of physics
 used a restricted arithmetic  and self reference did not produce
 paradoxical situations [340]. Or else the ultimate
 theory might actually be an infinite number
 of theories, but most having negligible effect, allows near certainty
 to be eventually attainable. In any
case, quantum mechanics alone seems
 dependent on observers making choices of possible measurements to
 take;
  and so argues against a single  objective history of  the
  universe [341]. Randomness would then play a more significant part in
  understanding the properties and values of the  various constants in the actual
  universe.

 Perhaps a single quanta of spacetime - a modern version of
Lema\^{i}tre's ``primeval atom''- is involved or an entirely new
conceptual approach ( from M theory?) is needed. The cosmological
aspects of quantum gravity certainly remains a fascinating topic
for much future work.

{\bf Acknowledgement}\\ This work developed from interesting email
discussions with Martin Bojowald. I should also like to thank
William Hiscock for remarks on the renormalization issue for CTCs.

\newpage

{\bf References}\\
\begin{enumerate}

\item J.J. Halliwell, in ``Quantum Cosmology and Baby Universes''
eds. S. Coleman, J.B. Hartle, T. Piran and S. Weinberg (World
Scientific: Singapore) 1991.
\item D.N. Page, in ``Proceedings of the Banff Summer institute on
Gravitation'', eds: R.B. Mann and P.S. Wesson (World Scientific:
Singapore) 1991.
\item D.L. Wiltshire, in ``Cosmology:
the Physics of the Universe'' eds. B. Robson, N. Visvanathon and
W.S. Woolcock (World Scientific: Singapore) 1996.
\item L.Z. Fang and Z.C. Wu, Int. J. of Mod. Phys. A 1 (1986)
p.887.
\item D. Atkatz, Am. J. Phys. 62 (1994) p.619.
\item J.B.  Hartle, in ``Proceedings of the 11th Nishinomiya-Yukawa
Symposium'', eds. K. Kikkawa, H. Kunitomo and H. Ohtsubo (World
Scientific: Singapore) 1998.
\item D.H. Coule, Phys. Rev. D 62 (2000) p.124010.
\item Z.C. Wu, ``No Boundary Universe'' (Hunan Sci. Tech Press
: Changsha) 1994.
\item A.O. Barvinsky, ``Talk at 9th Marcel Grossmann  Meeting'',
gr-qc/0101046.
\item F. Englert, presented at Erice School on ``Basics and Highlights in Fundamental
Physics'' 1999, hep-th/9911185.
\item T.P. Shestakova and C. Simeone, gr-qc/0409114 and
gr-qc/0409119.
\item K.V. Kuchar, ``Time and Interpretations of Quantum Gravity''
in ``Proceedings of the 4th Canadian Conference on General
Relativity and relativistic astrophysics'', eds. G. Kunstatter, D.
Vincent and J. Williams (World Scientific: Singapore) 1992.
\item M. Bojowald and H.A. Morales-Tecotl, Lect. Notes Phys. 646 (2004) p.421.\\
M. Bojowald, Gen. Rel. and Grav. 35 (2003) p.1877.\\
M. Bojowald, ``Talk at IUFM Marseille , France'' 2003 preprint
astro-ph/0309478.
\item T. Thiemann, Lect. Notes. Phys. 631 (2003) p.41.\\
A. Ashtekar and J. Lewandowski, Class. Quant. Grav. 21 (2004) p.R53;\\
A. Ashtekar, preprint math-ph/0202008;\\
A. Perez, Class. Quant. Grav. 20 (2003) p.R43.\\
for a simple introduction see also:\\ A. Ashtekar,  gr-qc/0410054.\\
C. Rovelli, Phys. World 16
(2003) p.37\\
L. Smolin, ``Three roads to quantum gravity'' (Oxford University
Press: Oxford) 2000.\\
some further text books are:\\
 R. Gambini and J. Pullin, ``Loops Knots, gauge theories and
quantum gravity'', ( Cambridge University Press : Cambridge)
1996.\\
 C. Rovelli, ``Quantum Gravity'', (Cambridge University Press :
Cambridge) 2004.\\
 T. Thiemann, ``Modern Canonical Quantum General
Relativity'' (Cambridge University Press: Cambridge) {\em To be
published}
\item L. Smolin, hep-th/0408048.\\
{\em ibid}, hep-th/0303185.
\item K. Kiefer, ``Quantum Gravity'' ( Oxford University Press:
Oxford) 2004.

\item ``Physical Origin of Time asymmetry'', eds. J.J. Halliwell,
J. P$\acute{e}$rez-Mercader W.H. Zurek (Cambridge University Press
: Cambridge) 1994.
\item H.D. Zeh, ``The Physical basis of the direction of time'',
(Springer-Verlag: Berlin) 2001.
\item ``The arguments of time'' ed. J. Butterfield ( Oxford
University Press: Oxford) 1999.
\item I. Prigogine, ``The End of Certainty'', (The Free Press:)
1997.
\item ``The Future of Theoretical Physics and
Cosmology'' eds. G.W. Gibbons, E.P.S. Shellard and S.J. Rankin
(Cambridge University Press: Cambridge) 2003.
\item ``Physics meets Philosophy at the Planck scale'', eds.
C. Callender and N. Huggett (Cambridge University Press:
Cambridge) 2001.
\item ``Science and Ultimate Reality'' eds. J.D. Barrow, P.C.W.
Davies and C.L. Harper (Cambridge University Press: Cambridge)
2004.
\item R. Penrose, ``The Road to Reality'', (Jonathon Cape: London)
2004.
\item R. Penrose, ``The Emperor's New Mind'', (Oxford University
Press: Oxford) 1989.
\item E.W. Kolb and M.S. Turner, ``The Early Universe''
(Addison-Wesley: Reading, MA) 1990.
\item A.D. Linde, ``Particle Physics and Inflationary Cosmology'',
(Harwood Academic Press: Chur, Switzerland) 1990.
\item S.W. Hawking and G.F.R. Ellis, ``The large scale structure
of space and time'' ( Cambridge University Press: Cambridge) 1973.
\item R.M. Wald, `` General Relativity'' (Chicago University
Press: Chicago) 1984.
\item M. Visser, ``Lorentzian Wormholes'' (AIP Press: New York)
1996.
\item E. Harrison, ``Cosmology 2nd ed.'' (Cambridge University Press:
Cambridge) 2000
\item J.V. Narlikar, ``An Introduction to Cosmology, 3rd. ed.'', (Cambridge
University Press: Cambridge) 2002.
\item J.D. Barrow and F.J. Tipler, ``The Anthropic Cosmological
Principle'', (Oxford University Press: Oxford) 1986.
\item W. Rindler, ``Relativity: Special, General and
Cosmological'' (Oxford university Press: Oxford) 2001.
\item Krasini\'{n}ski, ``Inhomogeneous Cosmological Models''
(Cambridge University Press: Cambridge) 1997.
\item S.M. Carroll, astro-ph/0310342.\\
T. Padmanabhan, gr-qc/0503107
\item T. Padmanabhan, ``Structure Formation in the Universe''
(Cambridge University Press: Cambridge) 1993 p.359.\\
see also: T. Padmanabhan and T.R. Seshadri, Class. Quant. Grav. 5
(1988) p.221.
\item L. Liu, F. Zhao and L.X. Li, Phys Rev. D 52 (1995) p.4752.
\item T. Padmanabhan and M.M. Vasanthi, Phys. Lett. A 89 (1982)
p.327.
\item O.M. Moreschi, gr-qc/9911105.\\
C. Schiller, gr-qc/9610066.
\item J.V. Narlikar and T. Padmanabhan, Phys. Rep. 100 (1983)
p.152.
\item S.W. Hawking and D.N. Page, Nucl. Phys. B 298 (1988) p.789.
\item G. Evrard and P. Coles, Class. Quant. Grav. 12 (1995) p.
L93.
\item D.H. Coule, Class. Quant. Grav. 13 (1996) p.2029.
\item A. Borde, A.H. Guth and A. Vilenkin, Phys. Rev. Lett. 90
(2003) p.151301.\\
for earlier proofs see:\\
 A. Borde and A. Vilenkin, in ``Proceedings of the Eighth Yukawa Symposium on Relativistic
Cosmology'' ed M. Sasaki (Universal Academy Press: Japan) 1995.\\
 A. Borde and A. Vilenkin, Int. J. Mod.
Phys. D 5 (1996) p.813.
\item E. Calabi and L. Marcus, Ann. Math. 75 (1962) p.63.\\
see also; H. J. Schmidt, Fortschr. Phys. 41 (1993) p.179.
\item A. Aguirre and S. Gratton, Phys. Rev. D 65 (2002) p.083507.
\item F. Hoyle and J.V. Narlikar, Proc. Roy. Soc. 277 (1964) p.1
\item P.J. Steinhardt and N. Turok, Phys. Rev. D 65 (2002)
p.126003.
\item J. Garriga and A. Vilenkin, Phys. Rev. D 57 (1998) p.2230.
\item Y.B. Zeldovich, ``My Universe'' (Harwood Academic Press)
1992 p.95.
\item V.F. Mukhanov, H.A. Feldman and R.H. Brandenberger,
Phys. Rept. 215 no. 5-6 (1992) p.203.
\item R.R. Caldwell, Phys. Lett. B 545 (2002) p.23.
\item F.G. Alvarenga and J.C. Fabris, Class. Quant. Grav. 13
(1996) p. L69.
\item D.H. Coule, Phys. Lett. B 450 (1999) p.48.
\item S.D.H. Hsu, A. Jenkins and M.B. Wise, Phys. Lett. B 597
(2004) p.270.
\item M.C.B. Abdalla, S. Nojiri and S.D. Odintsov,
hep-th/0409177.
\item S.M. Carroll, V. Duvvuri, M. Trodden and M.S. Turner,
Phys. Rev. D 70 (2004) p.043528.\\
 E. Bruning, D.H. Coule and C. Xu, Gen. Rel. Grav. 26 (1994)
p.1193.
\item P.J. Steinhardt and N. Turok, astro-ph/0404480.
\item C. Molina-Paris and M. Visser, Phys. Lett. B 455 (1999)
p.90.\\ B.K. Tippett and K. Lake, gr-qc/0409088.
\item R. Durrer and J. Laukenmann, Class. Quant. Grav. 13 (1996)
p.1069.
\item A.B. Batista, J.C. Fabris and S.V.B. Goncalves, Class.
Quant. Grav. 18 (2001) p.1389.
\item N. Pinto Neto, Int. J. Mod. Phys. D 13 (2004) p.1419.
\item J.D. Barrow and M.P. Dabrowski, Mon. Not. R. Astron. Soc.
275 (1995) p.850.
\item F. Hoyle, G. Burbidge and J.V. Narlikar, ``A Different
Approach to Cosmology'', (Cambridge University Press: Cambridge)
1999.
\item S.K. Banerjee and J.V. Narlikar, Astrophys. Journ. 487
(1997) p.69.
\item B.S. De Witt, Phys. Rev D 160 (1967) p.1113.
\item J.A. Wheeler, In `` Batelle Recontres'' eds. C. DeWitt and
J.A. Wheeler (Benjamin Press: New York) 1968.
\item P.C.W. Davies, New Scientist 5th March 2005 p.34
\item P. Ehrenfest, Proc. Amst. Acad. 20 (1917) p.200;\\
{\em ibid}, Ann. Physik 61 (1920) p. 440.
\item F.R. Tangherlini, Nuovo. Cim. 27 (1963) p.636
\item V.P. Frolov, M.A. Markov and V.F. Mukanov, Phys. Lett. B 216
(1989) p.272;\\
V. Mukhanov and R. Brandenberger, Phys. Rev. Lett. 68 (1992)
p.1969;\\
see also: M.A. Markov, Phys. Usp. 37 (1994) p.57.\\ V.P. Frolov
and I.D. Novikov, ``Black Hole Physics'' (Kluwer Academic
Publishers: Dordrecht) 1998.
\item L. Smolin, Class. Quant. Grav. 9 (1992) p.173.
\item L. Dyson, M. Kleban and L. Susskind, JHEP 0210 (2002)
p.011;\\
see also, D. Bak, hep-th/0208046.
\item A. Albrecht and L. Sorbo, Phys. Rev D 70 (2004) p.063528.
\item S.M. Carroll and J. Chen, hep-th/0410270.
\item G. 't Hooft, preprint gr-qc/9310026;\\
L. Susskind, J. Math. Phys. 36 (1995) p.6377;\\
for reviews of the holography principle see:\\
D. Bigatti and L. Susskind, hep-th/0002044.\\
L. Smolin, Nucl. Phys. B 601 (2001) p.209.
\item D.H. Coule, Int. J. Mod. Phys. D 12 (2003) p.963.
\item J.D. Barrow, `` Cargese Lectures in Physics'' eds. W.G.  Unruh
and J. Hartle ( Plenum Press: New York) 1987.
\item C.B. Collins and S.W. Hawking, Astrophys. J. 180 (1973)
p.317.
\item D.S. Goldwirth and T. Piran, Phys. Rep. 214 (1992) p.223.
\item W.G. Unruh, in ``Critical Dialogues in Cosmology'', ed. N.
Turok \\
( World Scientific: Singapore) 1997.
\item S.W. Hawking and R. Penrose, ``The Nature of Space and
Time''\\ ( Princeton University Press: Princeton) 1996 p.90.
\item M. Heusler, Phys. Lett. B 253 (1991) p.1991.
\item J. Ibanez and I. Olasagasti, Class. Quant. Grav. 15 (1998)
p.1937.
\item O. Iguchi and H. Ishihara, Phys. Rev. D 56 (1997) p.3216.
\item J.K. Erickson, D.H. Wesley, P.J. Steinhardt and N. Turok,
Phys. Rev. D 69 (2004) p.063514.
\item P.K.S. Dunsby, N. Goheer, M. Bruni and A. Coley,
Phys. Rev. D 69 (2004) p.R101303.
\item S. Hollands and R.M. Wald, Gen. Rel. Grav. 34 (2000) p.2043.
\item R. Penrose, in ``14th Texas symposium'', ed. E.J. Fenyves
(New York Academy of Sciences: New York) 1993.
\item R. Maartens, Living Rev. Rel. 7 (2004) p.1.
\item L. Randall and R. Sundrum, Phys. Rev. Lett. 83 (1999)
p.4690;\\ {\em ibid} p.3370.
\item M. Bucher, hep-th/0107148;\\
U.Gen, A. Ishibashi and T. Tanaka, Phys. Rev. D 66 (2002)
p.023519.
\item P. Anninos, R.A. Matzner and T. Rothman, Phys. Rev. D 43
(1991) p.3821.
\item C. Barcelo and M. Visser, Phys. Lett. B 482 (2000) p.183.
\item D.H. Coule, Class. Quant. Grav. 18 (2001) p.4265.
\item J.L. Hovdebo and R.C. Myers, JCAP 0311 (2003) p.012.
\item P. Kanti and K. Tamvakis, Phys. Rev. D 68 (2003) p.024014.
\item L.M. Burko and A. Ori, Phys. Rev. Lett. 74 (1995) p.1064.
\item M.G. Brown, K. Freese and W.H. Kinney, astro-ph/0405353.
\item R. Maartens, {\em private communication}
\item Y. Shtanov and V. Sahni, Phys. Lett. B 557 (2003) p.1.
\item G. Dvali, G. Gabadadze and G. Senjanovic, hep-ph/9910207.
\item M. Gasperini and G. Veneziano, Astropart. Phys. 1 (1993)
p.317.
\item M. Gasperini and G. Veneziano, Phys. Rept. 373 (2003) p.1.
\item D.H. Coule, Class. Quant. Grav. 15 (1998) p.2803.
\item A. Buonanno, T. Damour and G. Veneziano, Nucl. Phys. B 543
(1999) p.275.
\item Y.B. Zeldovich and I.D. Novikov, ``The
structure and evolution of the universe: relativistic astrophysics
vol.2'' (Chicago University Press: Chicago)1983 p.666.
\item D. Wands, Phys. Rev D 60 (1999) p.023507.
\item M.B. Miji\'{c}, Mod. Phys. Lett. A 12 (1997) p.647.
\item N.D. Birrell and P.C.W. Davies ``Quantum fields in curved
space'' (Cambridge University Press: Cambridge) 1982.
\item P. Martinetti and C. Rovelli, Class. Quant. Grav. 20 (2003)
p.4919.\\
S. Schlicht, Class. Quant. Grav. 21 (2004) p.4647.
\item S. Kawai, M. Sakagami and J. Soda, Phys. Lett. B 437 (1998)
p.284.
\item A.D. Linde, Phys. Lett. B 129 (1983) p.177.
\item A.D. Linde, Phys. Lett. B 175 (1986) p.395.
\item V.A. Belinsky, L.P. Grishchuk, Y.B. Zeldovich and I.M.
Klatatnikov, Sov. Phys. JEPT 62 (1985) p.195.\\ V.A. Belinsky and
I.M. Khalatnikov, Sov. Phys. JEPT 66 (1987) p.441.
\item M.S. Madsen and P. Coles, Nucl. Phys. B 298 (1988) p. 2757.
\item H.J. Schmidt, Astron. Nachr. 311 (1990) p. 99.
\item L. Kofman, A. Linde and V. Mukhanov, JHEP 0210 (2002) p.057.
\item G.W. Gibbons, S.W. Hawking and J.M. Stewart, Nucl. Phys. B
281 (1987) p.736.\\
M. Henneaux, Nuovo. Cimento Lett. 38 (1983) p.609.
\item D.N. Page, Phys. Rev. D 36 (1987) p.1607.
\item D.H. Coule, Class. Quant. Grav. 12 (1995) p.455.
\item S. Hollands and R.M. Wald, hep-th/0210001.
\item A. Vilenkin, Phys. Rev. Lett. 74 (1995) p.846.
\item D.H. Coule, Class. Quant. Grav. 10 (1993) p.L25.
\item S.W. Hawking, Nucl. Phys. B 239 (1984) p.257.
\item D.N. Page, Class. Quant. Grav. 1 (1984) p.417.\\
see also: A.Y. Kamenshchik, I.M. Khalatnikov and A.V. Toporensky,
Int. J. Mod. Phys. D 6 (1997) p.673.
\item N. Kanekar, V. Sahni and Y. Shtanov, Phys. Rev. D 63 (2001)
p.083520.
\item P. Chmielowski and D.N. Page, Phys. Rev. D 38 (1988) p.2392.
\item A. Linde, JCAP 0410 (2004) p.004.
\item A.D. Linde, Phys. Lett. B 327 (1994) p.208.\\
A. Vilenkin, Phys. Rev. Lett. 72 (1994) p.3137.
\item N. Sakai, Class. Quant. Grav. 21 (2004) p.281.
\item A.D. Linde, D.A. Linde and A. Mezhlumian, Phys. Rev. D 49
(1994) p.1783.
\item A.D. Linde and A. Mezhlumian, Phys. Rev. D 52 (1995) p.6789.
\item D.H. Coule, Gen. Rel. Grav. 36 (2004) p.2095.
\item  T.M. Davis, P.C.W. Davies and C.H. Lineweaver, Class. Quant.
Grav.20 (2003) p.2753.\\
P.C.W. Davies and T.M. Davis, Found. of Phys. 32 (2002) p.1877.
\item K. Kunze, Phys. Lett. B 587 (2003) p.1.
\item N. Turok, Class. Quant. Grav. 19 (2002) p.3449.
\item S. Blau, E. Guendelman and A.H. Guth, Phys. Rev. D 35 (1987)
p.1747.
\item E.H. Fahri and A.H. Guth, Phys. Lett. B 183 (1987) p.149.
\item K. Sato, M. Sasaki, H. Kodama and K. Maeda, Prog. Theor.
Phys. 65 (1981) p.143.\\
{\em ibid} Phys. Lett. B 108 (1982) p.103.
\item T. Vachaspati and M. Trodden, Phys. Rev. D 61 (2000) p.
023502.\\
{\em ibid}, Mod. Phys. Lett. A 14 (1999) p.1661.
\item J.M.M. Senovilla, Gen. Rel. Grav. 30 (1998) p.70.
\item D. Bak and S.J. Rey, Class. Quant. Grav. 17 (2000) p.L83.
\item R. Bousso, JHEP 9907 (1999) p.004.
\item N. Dadhich, J. Astrophysics. 18 (1997) p.343.
\item L.H. Ford, Proc. R. Soc. Lon. A 364 (1978) p.227.\\
L.H. Ford and T.A. Roman, Phys. Rev. D 55 (1997) p.2082.
\item E.H. Fahri, A.H. Guth and J. Guven, Nucl. Phys. B 339 (1990)
p.417.
\item W. Fischler, D. Morgan and J. Polchinski, Phys. Rev.
D 42 (1990) p.4042.\\
{\em ibid}, 42 (199) p.4042.
\item A.D. Linde, Nucl. Phys. B 372 (1992) p.421.
\item J.W. Moffat, Int. J. Mod. Phys. A 2 (1993) p.351.\\
A. Albrecht and J. Magueijo, Phys. Rev. D 59 (1999)
p.043516.\\
J.D. Barrow, Phys. Rev. D 59 (1999) p.043515.
\item J. Magueijo, Rept. Prog. Phys. 66 (2003) p.2025.
\item D.H. Coule, Mod. Phys. Lett. A 14 (1999) p.2437.
\item U.H. Gerlach, Phys. Rev. 177 (1969) p.1929.
\item I.V. Kanatchikov, Int. J. Theor. Phys. 40 (2001) p.1121.
\item J. B. Hartle in ref.[1]\\
M. Gell-Mann and J.B. Hartle, in Complexity, Entropy and the
Physics of Information, SFI studies in the science of Complexity
Vol. VIII, ed. W. Zurek (Addison Wesley) 1990.
\item E. Tryon, Nature 246 (1973) p.396.
\item R. Brout, F. Englert and E. Gunzig, Annalen der Physik, 115
(1978) p.78.
\item D. Atkatz and H. Pagels, Phys. Rev. D 25 (1982) p.2065.
\item A. Vilenkin, Phys. Rev. D 27 (1983) p.2848.
\item D.R. Brill, ``Quantum Cosmology'' ,  in `` Quantum theory and the structures of time and space: Feldafing
1974'' eds. L. Castell, M. Drieschner and C.F. von Weizsacker
(Carl Hanser Verlag: Munich ) p.231 1975
\item A.D. Linde, Lett. Nuovo Cim. 39 (1984) p.401.
\item A. Vilenkin, Phys. Rev. D 30 (1984) p.509.\\
A.D. Linde, Sov. Phys. JEPT 60 (1984) p.211.\\
V.A. Rubakov, Phys. Lett. B 148 (1984) p.280.
\item Y.B. Zeldovich
and A.A. Starobinsky, Sov. Astron. Lett. 10 (1984) p.135.
\item J.B. Hartle and S.W. Hawking, Phys. Rev. D 28 (1983) p.2960.
\item V.A. Rubakov and P.G. Tinyakov, Nucl. Phys. B 342 (1990)
p.430.
\item A. Vilenkin, Nucl. Phys. B 252 (1985) p.141.
\item A. Vilenkin, Phys. Rev. D 37 (1988) p.888.
\item A. Vilenkin, Phys. Rev. D 50 (1994) p.2581.
\item D.H. Coule, Mod. Phys. Lett. A 10 (1995) p.1989.
\item G.W. Gibbons and S.W. Hawking, Phys. Rev. D 15 (1977)
p.2752.\\
J. York, Phys. Rev. Lett. 28 (1972) p.1082.\\
see also: E. Poisson, ``A Relativist's Toolkit'' (Cambridge
University Press: Cambridge) 2004.
\item S.N. Solodukhin, Phys. Rev. D 62 (2000) p.044016.
\item E. Fahri, Phys. Lett. B 219 (1989) p.403.
\item G.W. Gibbons, S.W. Hawking and M.J. Perry, Nucl. Phys. B 138
(1978) p.141. \\
K. Schleich, Phys. Rev. D 36 (1987) p.2342.
\item E. Baum, Phys. Lett. B 133 (1983) p.185.
\item S.W. Hawking, Phys. Lett. B 134 (1983) p.403.
\item T. Padmanabhan, Phys. Lett. A 104 (1984) p.196.
\item E. Embacher, Class. Quant. Grav. 13 (1996) p.921;\\
{\em ibid}, ``Talk at Alexander Friedmann seminar,  St
Petersburg'' 1995 gr-qc/9507041
\item G.W. Gibbons, Class. Quant. Grav. 15 (1998) p.2605;\\
G.W. Gibbons, Nucl. Phys. B 472 (1996) p.683.
\item S.W. Hawking, Nucl. Phys. B 144 (1978) p.349.
\item J.P. Luminet, Phys. Reports  254 (1995) p.135;\\
Topology of the Universe conference, Clevelend 1997,  special
issue of Class. Quant. Grav. 15 (1998) p.2529.\\  for a simpler
introduction to compact manifolds in cosmology see:\\
 W.P. Thurston and J.R. Weeks, Sci. Amer.  251(7)  (1984) p.108;\\
J.P. Luminet,
G.D. Starkman and J.R.Weeks, Sci. Amer. April (1999).
\item S. Coleman, Nucl. Phys. B 310 (1988) p.643.
\item M. Duff, Phys. Lett. B 226 (1989) p.36.
\item A. Linde, Phys. Rev. D 58 (1998) p.083514.
\item H. Firouzjahi, S. Sarangi and S.H. Tye, JHEP 0409 (2004)
p.060.
\item R.W. Robinett, ``Quantum Mechanics'', (Oxford University
Press: Oxford) 1997.
\item J.J. Halliwell, Phys. Rev. D 39 (1989) p.2912.
\item C. Keifer, Class. Quant. Grav. 4 (1987) p.1369.\\
T. Padmanabhan, Phys. Rev. D 39 (1989) p.2924.\\
F. Fukuyama and M. Morikawa, Phys. Rev D 39 (1989) p.462.\\
 F. Mellor, Nucl. Phys. B 353 (1991) p.291.\\
 A.O. Barvinsky, A.Y. Kamenshchik, C. Kiefer and I.V. Mishakov,
 Nucl. Phys. B 551 (1999) p.374.
\item E. Calzetta and E. Verdaguer, Phys. Rev. D 59 (1999)
p.083513.
\item J.M. Cline, Phys. Lett. B 224 (1989) p.53;\\
M.B. Miji\'{c}, Phys. Lett. B 241 (1990) p.242;\\
A. Strominger, Nucl. Phys. B 319 (1989) p.722.
\item D. Marolf, Phys. Rev. D 53 (1996) p.6979.
\item O.Y. Shvedov, Surveys High Energy Phys. 10 (1997) p.411.
\item H. Fink and H. Leschke, Found. Phys. Lett. 13 (2000) p.345.
\item D.H. Coule and J. Martin, Phys. Rev. D 61 (2000) p.063501.
\item D.H. Coule, Mod. Phys. Lett. A 13 (1998) p.961.
\item Computer software, MAPLE V release 4.6, Waterloo Maple Inc.
1996.
\item M. Abramowitz and I.A. Stegun, ``Handbook of Mathematical
Functions'' (Dover Press) 1965.
\item S.W. Hawking and D.N. Page, Phys. Rev. D 42 (1990) p.2655.
\item C.M. Bender and S.A. Orszag, ``Advanced Mathematical Methods
for Scientists and Engineers'' (McGraw-Hill: ) 1984.
\item S.W. Hawking and D.N. Page, Nucl. Phys. B 264 (1986) p.185.
\item S.P. Kim, Phys. Rev. D 46 (1992) p.3403.
\item N. Kontoleon and D.L. Wiltshire, Phys. Rev. D 59 (1999)
p.063513.
\item N.A. Lemos, J. Math. Phys. 37 (1996) p.1449.\\
J.H. Kung, Gen. Rel. Grav. 27 (1995) p.35.
\item J. Feinberg and Y. Peleg, Phys. Rev. D 52 (1995) p.1988.
\item D.H. Coule, Class. Quant. Grav. 9 (1992) p.2353.\\
A.K. Sanyal, Int. J. Mod. Phys. A 10 (1995) p.2231.
\item G. Magnano and L.M. Sokolowski, Phys. Rev. D 50 (1994)
p.5039.
\item A. Carlini, D.H. Coule and D.M. Solomons, Mod. Phys. Lett. A
18 (1996) p.1453.
\item A. Carlini, D.H. Coule and D.M. Solomons, Int. J. of Mod.
Phys. A 12 (1997) p.3517.
\item J.B. Hartle, J. Math. Phys. 30 (1989) p.452.
\item W.M. Suen and K. Young, Phys. Rev. D 39 (1989) p.2201.
\item D.N. Page, Phys. Rev. D 56 (1997) p.2065.
\item V.A. Rubakov, Phys. Lett. B 148 (1984) p.280.\\
V.A. Rubakov, gr-qc/9910025.\\
 V.T. Gurovich, H.J. Schmidt and
I.V. Tokareva, Gen. Rel. Grav. 33
(2001) p.591.\\
 J. Hong, A. Vilenkin and S. Winitzki, Phys. Rev. D
68 (2003) p.023520.\\
S.P. Kim, gr-qc/0403015
\item A. Vilenkin, Phys. Rev. D 58 (1998) p.067301.\\
A. Vilenkin, gr-qc/9812027.\\
R. Bousso and S.W. Hawking, Grav. Cosmol. Suppl. 4 (1998) p.28.
\item L.P. Grishchuk and Y.V. Sidorov, Sov. Phys. JEPT 67 (1988)
p.1533.
\item G.W. Gibbons and L.P. Grishchuk, Nucl. Phys. B 313 (1989)
p.736.
\item L.P. Grischuk, Class. Quant. Grav. 10 (1993) p.2449.
\item P.D. D'Eath, ``Supersymmetric Quantum Cosmology'' (Cambridge
University Press: Cambridge) 1995.
\item J.R. Gott and Li-Xin Li, Phys. Rev. D 58 (1998) p.023501.
\item G.W. Gibbons and J.B. Hartle, Phys. Rev. D 42 (1990) p.2458.
\item D. Green and W.G. Unruh, preprint gr-qc/0206068.
\item N. Pinto-Neto, Found. Phys. 35 (2005) p.577.
\item D.N. Page, gr-qc/0001001.
\item S.B. Giddings and A. Strominger, Nucl. Phys. B 321 (1989)
p.481.\\
M. McGuigan, Phys. Rev. D 38 (1988) p.3031.
\item R. Geroch and J.B. Hartle, Found. of Phys. 16 (1986) p.533.
\item J.B. Hartle, in ``Boundaries and Barriers: On the limits of
Scientific Knowledge'' eds: J.L. Casti and A. Karlqvist
(Addison-Wesley: Reading MA) 1996.
\item K. Schleich and D.M.  Witt, Phys. Rev. D 60 (1999)
p.064013.\\
J.B. Hartle, Class. Quant. Grav. 2 (1985) p.707.
\item S. Carlip, Phys. Rev Lett. 79 (1998) p.4071.\\
S. Carlip, Class. Quant. Grav. 15 (1998) p.2629.
\item M. Anderson, S. Carlip, J.G. Ratcliffe, S. Surya and S.T.
Tschantz, Class. Quant. Grav. 21 (2004) p.729.\\
see also:  Y. Fujiwara, S. Higuchi, A. Hosoya, T. Mishima and M.
Siino, Phys. Rev. D 44 (1991) p.1756.
\item G.W. Gibbons, in ref.[21].
\item J. Stelmach and I. Jakacka, Class. Quant. Grav. 18 (2001)
p.2643.
\item J. Yokoyama and K. Maeda, Phys. Rev. D 41 (1990) p.1047.
\item V.M. Mostepanenko and N.N. Trunov, ``The Casimir effect and
its applications'' (Oxford University Press: Oxford) 1997.\\
see also:  D. Muller, gr-qc/0403086.
\item C.J. Isham, Proc. R. Soc. A 362 (1978) p.383.
\item J.A. Belinchon, Int. J. Mod. Phys. D 11 (2002) p.527.
\item R. Mansouri and F. Nasseri, Phys. Rev. D 60 (1999) p.123512.
\item H. Conradi, Int. J. Mod. Phys. D 7 (1998) p.189.
\item J.J. Halliwell and S.W. Hawking, Phys. Rev. D 31 (1985)
p.1777.
\item S.W. Hawking, R. Laflamme and G.W. Lyons, Phys. Rev. D 47
(1993) p.5342.
\item T. Vachaspati and A. Vilenkin, Phys. Lett. B 217 (1989)
p.228.
\item E.W. Kolb, S. Matarrese, A. Notari and A. Riotto,
hep-th/0503117.
\item T. Tanaka and M. Sakagami, Prog. Theo. Phys. 100 (1998)
p.547.\\
A.L. Matacz, Phys. Rev. D 49 (1994) p.788.\\
M. Sakagami, Prog. Theo. Phys. 79 (1988) p.442.
\item D.N. Page, Phys. Rev. D 32 (1985) p.2496
\item S.W. Hawking, in ref.[17]
\item J.J. Halliwell, in ref.[17]
\item C. Kiefer and H.D. Zeh, Phys. Rev. D 51 (1995) p.4145.
\item C. Kiefer, Phys. Rev. D 38 (1988) p.1761.
\item C. Kiefer, gr-qc/0502016.
\item S.W. Hawking, Phys. Scr. T 15 (1987) p.151.\\
D.N. Page, Phys. Rev. D 34 (1986) p.2267.
\item M. Gell-Mann and J.B. Hartle, in ref.[17]
\item P.C.W. Davies and J. Twamley, Class. Quant. Grav. 10 (1993)
p.931.
\item J. Garriga and M. Sasaki, Phys. Rev. D 62 (2000) p.043523.\\
S. Nojiri, S.D. Odintsov and S. Zerbini, Class. Quant. Grav. 17
(2000) p.4855.\\ A.S. Gorsky and K.G. Selivanov, Phys. Lett. B 485
(2000) p.271. \\ A.S. Gorsky and K.G. Selivanov,
 Int. J. of Mod. Phys. A 16 (2001) p.2243.\\
K. Koyama and J. Soda, Phys. Lett. B 483 (2000) p.043501. \\ L.
Anchordoqui, C. Nunez and K. Olsen,
JHEP 0010 (2000) p.050.\\
 S.W. Hawking, T. Hertog and H.S. Reall,
Phys. Rev D 63 (2001) p.083504.\\
M. Bouhmadi-Lopez, P.F. Gonzalez-Diaz and A. Zhuk, Class. Quant.
Grav. 19 (2002) p.4863.\\ S.S. Seahra, H.R. Sepangi and J. Ponce
de Leon, Phys. Rev. D 68 (2003) p.066009.\\ R. Cordero and E.
Rojas, Class. Quant. Grav. 21 (2004) p.4231.\\
Z.C. Wu, hep-th/0405249.\\
 A. Boyarsky, A. Neronov and I. Tkachev,
gr-qc/0411144.
\item R. Cordero and A. Vilenkin, Phys. Rev. D 65 (2002) p.083519.
\item E.I. Guendelman and A.B. Kaganovich, Grav. Cosmol. 1 (1995) p.103.\\
A. Davidson, Class. Quant. Grav. 16 (1999) p.653. \\ A. Davidson,
D. Karasik and Y. Lederer, Class. Quant. Grav. 16 (1999) p. 1349.\\
F. Darabi, W.N.
Sajko and P.S. Wesson, Class. Quant. Grav. 17 (2000) p.4357.\\
P.I. Fomin and Yu.V. Shtanov, Class. Quant. Grav. 19 (2002)
p.3139.
\item E. Carugno, M. Litterio, F. Occhionero and G. Pollifrone,
Phys. Rev. D 53 (1996) p.6863.
\item K. Aoyanagi and K. Maeda, Phys. Rev. D 70 (2004) p.123506.
\item S. Nojiri and S.D. Odintov, hep-th/0409244.
\item R. Gambini, R.A. Porto and J. Pullin, Phys. Rev. Lett. 93 (2004) p.240401.\\
{\em ibid}, Int. J. of Mod. Phys. D 13 (2004) p.2315.\\
 {\em ibid}, New J. Phys. 6 (2004) p.45.
\item S.W. Hawking, ``Talk at GR17 Dublin'' 2004
\item Li-Xin Li and J.R. Gott, Phys. Rev. Lett. 80 (1998) p.2980.
\item W.A. Hiscock and D.A. Konkowski, Phys. Rev. D 26 (1982)
p.1225.
\item P.F. Gonz\'{a}lez-D\'{i}az, Phys. Rev. D 59 (1999)
p.123513.
\item J.S. Dowker, Phys. Rev. D 18 (1978) p.1856.
\item S.W, Hawking, Commun. Math. Phys. 43 (1975) p.199.
\item G.W. Gibbons and S.W. Hawking, Phys. Rev. D 15 (1977)
p.2738.
\item W.A. Hiscock, preprint gr-qc/0009061.
\item Li-Xin Li, Phys. Rev. D 59 (1999) p.084016.
\item V.L. Ginzburg and V.P. Frolov, Usp. Fiz. Nauk 150 (1986) p.4.\\
L.P. Grishchuk, Y.B. Zeldovich and L.V. Rozhanski, Sov. Phys. JEPT
65 (1987) p.11.
\item P.R. Anderson, W.A. Hiscock and B.E. Taylor, Phys. Rev. Lett.
85 (2000) p.2438.
\item G.F.R. Ellis and R. Maartens, Class. Quant. Grav. 21 (2004)
p.223.
\item J.L. Friedman and A. Higuchi, Phys. Rev. D 52 (1995) p.5687.
\item J.D. Barrow, G.F.R. Ellis, R. Maartens and C.G. Tsagas,
Class. Quant. Grav. 20 (2003) p.L155.
\item H.D. Conradi and H.D. Zeh, Phys. Lett. A 154 (1991) p.321.\\
H.D. Conradi, Phys. Rev. D 46 (1992) p.612.
\item M.B. Miji\'{c}, M.S. Morris and W. Suen, Phys. Rev. D 39
(1989) p.1496.
\item R. Coliste, J.C. Fabris and N. Pinto-Neto, Phys. Rev. D 62
(2000) p.83507.
\item F.G. Alvarenga, J.C. Fabris, N.A. Lemos and G.A. Monerat,
Gen. Rel. Grav. 34 (2002) p.651.
\item G.F.R. Ellis, A. Sumeruk, D.H. Coule and C. Hellaby, Class.
Quant. Grav. 9 (1992) p.1535.\\
G.F.R. Ellis, Gen. Rel. Grav. 24 (1992) p.1047.
\item F. Embacher, Phys. Rev. D 51 (1995) p.6764.
\item F. Embacher, Phys. Rev. D 52 (1995) p.2150.
\item R. Coliste, J.C. Fabris and N. Pinto-Neto, Phys. Rev. D 57
(1998) p.4707.
\item S.W. Hawking, Phys. Rev. D 14 (1976) p.2460.
\item J.G. Russo, hep-th/0501132
\item G.T. Horowitz and J. Maldacena, JHEP 0402 (2004) p.008.
\item H. Weinfurter, Phys. World, Jan 2005 p.47.
\item P.M. Alsing, D. McMahon and G.J. Milburn, quant-ph/0311096.
\item D. Gottesman and J. Preskill, JHEP 0403 (2004) p.026.
\item S. Lloyd, quant-ph/0406205
\item S. Lloyd, Phys. Rev. Lett. 88 (2001) p.237901.
\item U. Yurtsever and G. Hockney, Class. Quant. Grav. 22 92005)
p.295.
\item H. Kodama, Phys. Rev. D 42 (1990) p.2548.
\item C. Soo, Class. Quant. Grav. 19 (2002) p.1051.\\
E. Witten, gr-qc/0306083.\\
 L. Freidel and L. Smolin, Class.
Quant. Grav. 21 (2004) p.3831.
\item  S. Alexander, J. Malecki and L. Smolin, Phys. Rev. D 70
(2004) p.044025.
\item V. Husain and O. Winkler, Phys. Rev. D 69 (2004) p.084016.
\item G.F.R. Ellis and D.R. Matravers, Gen. Rel. Grav. 27 (1995) p.777.\\
R. Zalaletdinov, R. Tavakol and G.F.R. Ellis, Gen. Rel.
Grav. 28 (1996) p.1251.
\item W. Rindler, Phys. Lett. A 276 (2000) p.52.
\item M. Bojowald, Class. Quant. Grav. 19 (2002) p.5113.
\item M. Bojowald and F. Hinterleitner, Phys. Rev. D 66 (2002)
p.104003.
\item M. Bojowald, Class. Quant. Grav. 18 (2001) p.L109.
\item M. Bojowald, Phys. Rev. Lett. 89 (2002) p.261301.
\item M. Bojowald, Phys. Rev. Lett. 86 (2001) p.5227.
\item M. Bojowald and K. Vandersloot, Phys. Rev. D 67 (2003)
p.124023.
\item M. Bojowald and G. Date, Phys. Rev. Lett. 92 (2004)
p.071302.\\
M. Bojowald, G. Date and G.M. Hossain, Class. Quant. Grav. 21
(2004) p.3541.
\item J.D. Barrow, Nature 272 (1978) p.211.
\item J.D. Barrow, Phys. Rept. 85 (1982) p.1.
\item D. Green and W.G. Unruh, Phys. Rev. D 70 (2004) p.103502.
\item J. Kowalski-Glikman and J.C. Vink, Class. Quant. Grav. 7
(1990) p.901.
\item S. Mollerach, S. Matarrese and F. Lucchin, Phys. Rev. D 50
(1994) p.4835.
\item V.A. Rubakov, M.V. Sazhin and A.V. Veryaskin, Phys. Lett. B
115 (1982) p.189.\\
L.F. Abbott and M.B. Wise, Nucl. Phys. B 244 (1984) p.541.
\item T. Padmanabhan, Phys. Rev. Lett. 60 (1988) p.2229.\\
T. Padmanabhan, T.R. Seshardri and T.P. Singh, Phys. Rev. D 39
(1989) p.2100.
\item S. Tsujikawa, P. Singh and R. Maartens, Class. Quant. Grav.
21 (2004) p.5767.
\item M. Bojowald, J.E. Lidsey, D.J. Mulryne, P. Singh and R.
Tavakol, Phys. Rev. D 70 (2004) p.043530.
\item G.M. Hossain, gr-qc/0503065
\item D.H. Coule, gr-qc/0312045.
\item G.V. Vereshchagin, JCAP 0407 (2004) p.013.
\item M. Bojowald, R. Maartens and P. Singh, Phys. Rev. D 70
(2004) p.083517.
\item G. Date and G.M. Hossain, Phys. Rev. Lett. 94 (2005)
p.011302.
\item D.J. Mulryne, R. Tavakol, J.E. Lidsey and G.F.R. Ellis,
astro-ph/0502589.
\item P. Singh, gr-qc/0502086.
\item R. Graham and P. Szepfalusy, Phys. Rev. D 42 (1990) p.2483.
\item A.D. Linde, {\em private communication}
\item M. Bojowald, R. Goswami, R. Maartens and P. Singh,
gr-qc/0503041.
\item Y.J. Ng and H. van Dam, Found. Phys. 30 (2000) p.795.
\item G. Amelino-Camelia, Mod. Phys. Lett. A 17 (2002) p.899.
\item C. Simon and D. Jaksch, quant-ph/0406007.
\item ``Inhomogeneous cosmological models'' eds. A. Molina and
J.M.M. Senovilla (World Scientific Press: Singapore) 1995
\item L.H. Ford, gr-qc/0504096
\item T. Padmanabhan and T.R. Choudhury, Mod. Phys. Lett. A 15
(2000) p.1813.
\item N. Ohta, Int. J. Mod. Phys. A 20 (2005) p.1.
\item P.C.W. Davies, ``When Time Began'', New Scientist (special supplement) 9/10/2004.
\item G.F.R. Ellis, in ``Modern Cosmology'', eds: S. Bonometto,
V. Gorini and U. Moshella, (IOP Press: Bristol) 2001.\\
G.F.R. Ellis, Class. Quant. Grav. 16 (1999) p.A37.
\item G. 't Hooft, quant-ph/0212095.
\item W.H. Zurek, Phys. Rev. A 40 (1989) p.4731.
\item J.D. Barrow, New. Astron. 4 (1999) p.333.\\
see also: P.C.W. Davies in ref.[17].
\item A.A. Grib and Y.V. Pavlov, Grav. Cosmol. Suppl. 8N1 (2002) p.148: preprint gr-qc/0206040.
\item B. McInnes, Nucl. Phys. B 709 (2005) p.213.
\item L. Susskind, hep-th/0302219.\\
M. Dine, hep-th/0402101.
\item G.L. Kane, M.J. Perry and A.N. Zytkow, New. Astron. 7 (2002)
p.45.
\item B.R. Frieden, ``Physics from Fischer Information-a
unification'' (Cambridge University Press: Cambridge) 1998; \\
see also: H. Christian von Baeyer, `` Information the new Language
of Science'' (Weidenfeld and Nicolson: London) 2003.\\
M. Gell-Mann, ``The Quark and the Jaguar'' (Little, Brown and
Company: London) 1994.
\item V. Dzhunushaliev and D. Singleton, Entropy 4 (2002) p.2.
\item S.W. Hawking, Lecture `` Godel and the End of Physics''
available at: www.damtp.cam.ac.uk/strtst/dirac/hawking/
\item J.D. Barrow, ``Impossibility'' (Oxford University Press:
Oxford) 1998.
\item S.W. Hawking, Lecture at KITP, Santa Barbara, USA (2003)

\end{enumerate}
\end{document}